\documentclass[journal,12pt,draftclsnofoot,onecolumn,english]{IEEEtran}

\usepackage{epsfig}
\usepackage{times}
\usepackage{float}
\usepackage{afterpage}
\usepackage{sansmath}
\usepackage{amsmath}
\usepackage{amstext,cite}
\usepackage{amssymb,bm}
\usepackage{latexsym}
\usepackage{color}
\usepackage{graphicx}
\usepackage{amsmath}
\usepackage{amsthm}
\usepackage{graphicx}
\usepackage[center]{caption}
\usepackage{pstricks}
\usepackage{caption}
\usepackage{subcaption}
\usepackage{booktabs}
\usepackage{multicol}
\usepackage{lipsum}
\usepackage[T1]{fontenc}
\usepackage{hyperref}
\usepackage{aecompl}
\usepackage{mathrsfs}
\usepackage{blkarray}

\usepackage{arydshln}

\usepackage{soul}
\usepackage{cancel}

\usepackage{changes}
\definechangesauthor[color=red,name={Mingyue Ji}]{MJ}

\allowdisplaybreaks

\long\def\comment#1{}

\newcommand{\av}{{\mathbf a}}

\newcommand{\ev}{{\mathbf e}}

\newcommand{\sv}{{\mathbf s}}

\newcommand{\Ac}{{\mathcal A}}
\newcommand{\Bc}{{\mathcal B}}
\newcommand{\Cc}{{\mathcal C}}

\newcommand{\Oc}{{\mathcal O}}

\newcommand{\Rc}{{\mathcal R}}
\newcommand{\Sc}{{\mathcal S}}

\newcommand{\Uc}{{\mathcal U}}

\newcommand{\Vc}{{\mathcal V}}
\newcommand{\Xc}{{\mathcal X}}

\newcommand{\Zc}{{\mathcal Z}}

\newcommand{\qsf}{{\mathsf q}}

\newcommand{\Ksf}{{\mathsf K}}
\newcommand{\Lsf}{{\mathsf L}}

\newcommand{\Rsf}{{\mathsf R}}
\newcommand{\Ssf}{{\mathsf S}}
\newcommand{\Tsf}{{\mathsf T}}
\newcommand{\Usf}{{\mathsf U}}

\newtheorem{thm}{Theorem}

\newtheorem{lem}{Lemma}

\newtheorem{rem}{Remark}
\newtheorem{example}{Example}

\providecommand{\definitionname}{Definition}

\usepackage{amsmath}
\usepackage{tikz}
\usetikzlibrary{calc}

\usepackage[nodisplayskipstretch]{setspace} 

\begin{document}

\title{The Capacity Region of Information Theoretic Secure Aggregation with Uncoded Groupwise Keys}  
\author{
Kai~Wan,~\IEEEmembership{Member,~IEEE,} 
Hua~Sun,~\IEEEmembership{Member,~IEEE,}
Mingyue~Ji,~\IEEEmembership{Member,~IEEE,}  
Tiebin~Mi,~\IEEEmembership{Member,~IEEE,}  
and~Giuseppe Caire,~\IEEEmembership{Fellow,~IEEE}
\thanks{
K.~Wan and T.~Mi are with the School of Electronic Information and Communications,
Huazhong University of Science and Technology, 430074  Wuhan, China,  (e-mail: \{kai\_wan,mitiebin\}@hust.edu.cn). The work of K.~Wan and  T.~Mi was partially funded by the   National Natural
Science Foundation of China (NSFC-12141107).}
\thanks{
M.~Ji is with the Electrical and Computer Engineering Department, University of Utah, Salt Lake City, UT 84112, USA (e-mail: mingyue.ji@utah.edu). The work of  M.~Ji was supported in part by NSF CAREER Award 2145835 and NSF Award 231222.}
\thanks{
H.~Sun is with the Department of Electrical Engineering, University of North Texas, Denton, TX 76203, USA (email: hua.sun@unt.edu). The work of H. Sun was supported in part by NSF Awards 2007108 and 2045656.
}
\thanks{G.~Caire is with the Electrical Engineering and Computer Science Department, Technische Universit\"at Berlin, 10587 Berlin, Germany (e-mail: caire@tu-berlin.de). The work of  G.~Caire was partially funded by the European Research Council under the ERC Advanced Grant N. 789190, CARENET.}
}
\maketitle

\begin{abstract}
This paper considers the secure aggregation problem for federated learning under an information theoretic cryptographic formulation, where distributed training nodes (referred to as users) train models based on their own local data and a curious-but-honest server aggregates the trained models without retrieving other information about users' local data.
Secure aggregation generally contains two phases, namely key sharing phase and model aggregation phase. 
Due to the common  effect of user dropouts in federated learning, the model aggregation phase should contain two rounds, where in the first round the users transmit masked models and, in the second round, according to 
  the identity of surviving users after the first round, these surviving users  transmit some further messages to help the server decrypt the sum of users' trained models. 
The objective of the considered information theoretic formulation  is to characterize the capacity region of the communication rates in the two rounds from the users to the server in the model aggregation phase,  assuming that  key sharing has already been performed offline in prior. 
In this context, Zhao and Sun completely characterized the capacity region 
under the assumption that the keys can be arbitrary random variables. More recently, an additional constraint, known as ``uncoded groupwise keys,'' has been introduced. This constraint entails the presence of multiple independent keys within the system, with each key being shared by precisely $\Ssf$ users, where $\Ssf$ is a defined system parameter.
The capacity region for the information-theoretic secure aggregation problem with uncoded groupwise keys was established in our recent work
 subject to the condition $\Ssf>\Ksf-\Usf$,  where $\Ksf$ is the number of total users and $\Usf$ is the designed minimum number of surviving users (which is another system parameter).  In this paper we fully characterize of the the capacity region for this problem by proposing a new converse bound and an achievable scheme. Experimental results over the Tencent Cloud show the improvement on the model aggregation time compared to the original secure aggregation scheme.
\end{abstract}


\section{Introduction}
\label{sec:intro}
\subsection{Background on secure aggregation for federated learning}
\label{sub:background}
Federated learning is a decentralized machine learning approach that enables multiple devices or users (such as smartphones, edge devices, or Internet of Things devices) to collaboratively train a shared model without sharing their local raw data to the central server~\cite{mcmahan2017communication,yang2019federated,li2020federated,mcmahan2021advances}. Rather than centralizing all data in a single location, federated learning allows each device training by using its own local data.
After initialization, the process of federated learning    involves several iterations among the users and the server. In one iteration,   each user trains the model using its own local data without sharing it with the central server. After training on local data, the users send their model updates (weights or gradients) to the   server.
Then the central server collects the model updates from all the users and aggregates the updated models 
to create an updated global model.
Federated learning has two main advantages over traditional centralized and distributed learning: (i) it reduces communication costs and eliminates the need for frequent data transfers; (ii) it preserves data privacy against the server by keeping data local. 
Despite these advantages,   federated learning also suffers from some challenges. On the one hand,  assume that the  training devices/users are  smartphones or  edge devices; during the training process of federated learning, the server may lose the connectivity to some users due to user mobility and fluctuating communication quality. Thus an efficient federated learning scenario should be  resilient to this unpredictable effect of user dropouts. On the other hand,   each user  needs to transmit to the server the computed model in terms of the local data; thus the information of local data can be leaked at some level to the server, and this is known as the model inversion attacks in federated learning~\cite{inverting2020geiping}.

  To deal with the effect of user dropouts and strengthen local data privacy in federated learning, 
a new cryptographic problem, referred to as secure aggregation,  was originally introduced in~\cite{bonawitz2017practical}. 
Except the desired sum of the users' updated models, the server should not learn other information about the users' local data. 
In order to guarantee the computational or information theoretic security, 
the key-based encryption could be used, where keys are  shared among the users and thus the users' computed updated models (e.g., sub-gradients) could be masked by the keys. 
The keys are generated and then  shared  to the users according to some key generation protocols. If the key generation is independent of the training data, the key sharing is called offline; otherwise, it is called online. 
Model aggregation follows key sharing, where the users compute, mask, and send their computed updated models to the server. For the resilience  to user dropouts, multi-round communications among the users and server could be used, where in the first round the users send masked updated models and in the remaining rounds the users send some messages composed of keys.  
 The number of rounds depend on the threat models of the server (e.g., the server may be honest-but-curious, or malicious and lie on the identity of the dropping users, or even collude with some users).
The secure aggregation protocol in~\cite{bonawitz2017practical} uses the pairwise offline key sharing (i.e., each pair of users share a key) based on Shamir's secret sharing~\cite{share1979Shamir}  in order to deal with user dropouts.

Following the secure aggregation problem with user dropouts in~\cite{bonawitz2017practical}, several works have developed more efficient and/or more secure schemes for  aggregation,
 for example, by using common seeds   through  homomorphic pseudorandom generator~\cite{efficientDropout}, secure multi-party computing~\cite{hetero2022Elkordy},  non-pairwise keys~\cite{ITsecureaggre2021}, online key sharing~\cite{so2021turbo,kadhe2020fastsecagg,nezhad2022swiftagg}, improved El Gamal encryption~\cite{yang2023effcient}. 
  The readers  can refer to the survey for more details~\cite{bigdatareview,elkordy2023federatedana}.

\subsection{Information theoretic secure aggregation}
\label{sub:IT secure}
In this paper,  we follow the $(\Ksf,\Usf )$ information theoretic formulation on  secure aggregation  with   user dropouts and offline key sharing proposed in~\cite{ITsecureaggre2021}, where $\Ksf$ represents the number of users in the system and $\Usf$ represents the minimum number of non-dropped users. The input vector (i.e., updated model) of each user $k$ is denoted by $W_k$, which contains   $\Lsf$ uniform and i.i.d. symbols over a  finite field.\footnote{\label{foot:design U}
Note that  the system is designed to tolerate up to $\Ksf - \Usf$ user dropouts. If  more than $\Ksf - \Usf$ users  drop, there is insufficient data for update and the server does not update the model and will ask for a retransmission. In this paper, we directly assume that there are at most $\Ksf - \Usf$ user dropouts. 
}
The main information theoretic problem for this model is to determine the optimal transmission   in the model aggregation phase with the assumption that enough keys  have been shared among the users in a prior key sharing phase, such that the information theoretic security of the users’ local data is protected against the server (with the exception of the sum of the updated models of the non-dropped users).\footnote{\label{foot:inform theory security}
Information theoretic security was proposed in the seminal work by Shannon~\cite{shannonsecurity}, under which constraint even if the  adversary has infinite computation power it still cannot get any information about the data. 
In the literature of secure aggregation with user dropouts, the secure aggregation schemes proposed in~\cite{ITsecureaggre2021,lightsec2021so,nezhad2022swiftagg} guarantee the   information theoretic security constraint.}
Thus each user $k$ has a key $Z_k$, which can be any random variable    independent of  $W_1,\ldots,W_{\Ksf}$.
It was proved in~\cite{ITsecureaggre2021} that to preserve the security of users' local data against the honest-but-curious server with the existence of user dropouts,  two-round transmission in the model aggregation phase is necessary and also sufficient.  
In the first round,  each user  masks its input vector by the stored key and transmits the masked input vector to the server. The server receives and then returns a feedback to the non-dropped users about the identity of the non-dropped users. In the second round, each   non-dropped user  further transmits a coded message as a function of its local data, key, and 
  the server feedback. The users may also drop in the second round; the secure aggregation scheme should guarantee that    by the two-round transmission the server could recover the  sum of the input vectors of the non-dropped users in the first  round. 
The objective of this problem is to characterize the region of all possible achievable   rate tuples $(\Rsf_1,\Rsf_2)$, where $\Rsf_i$ represents the largest number of transmissions in the $i^{\text{th}}$ transmission round among all users normalized by $\Lsf$ for $i\in \{1,2\}$.  
The capacity region was proved to be $\{(\Rsf_1,\Rsf_2):\Rsf_1 \geq 1,\Rsf_2 \geq 1/\Usf \}$ in~\cite{ITsecureaggre2021} with an achievability strategy based on Minimum Distance Separable (MDS) codes in the key generation and one-time pad coding in the  model aggregation. Another secure aggregation scheme which can also achieve   capacity was proposed in~\cite{lightsec2021so}, based on a pairwise coded key generation. Compared to~\cite{ITsecureaggre2021}, the scheme in~\cite{lightsec2021so} significantly reduces the size of keys stored by each user.

There are  some other extended information theoretic formulations on secure aggregation in federated learning, following the model in~\cite{ITsecureaggre2021}. 
A weaker information theoretic  security constraint compared to the one in~\cite{ITsecureaggre2021} was considered in~\cite{zhou2023weekly}, where  we only need to preserve the security on a subset of users' input vectors. 
User collusion  was also  considered in~\cite{ITsecureaggre2021}, where the server may collude with up to $\Tsf<\Usf$ users; to deal with potential user collusion, the capacity region reduces to $\{(\Rsf_1,\Rsf_2):\Rsf_1 \geq 1,\Rsf_2 \geq 1/(\Usf-\Tsf) \}$,   characterized in~\cite{ITsecureaggre2021}.  Information theoretic secure aggregation with cluster federated learning was originally considered in~\cite{sami2023cluster},  to aggregate the updated models from multiple clusters of users simultaneously, without learning any information about the cluster identities or users' local data. 

\begin{figure}
    \centering
    \begin{subfigure}[t]{1\textwidth}
        \centering
        \includegraphics[scale=0.25]{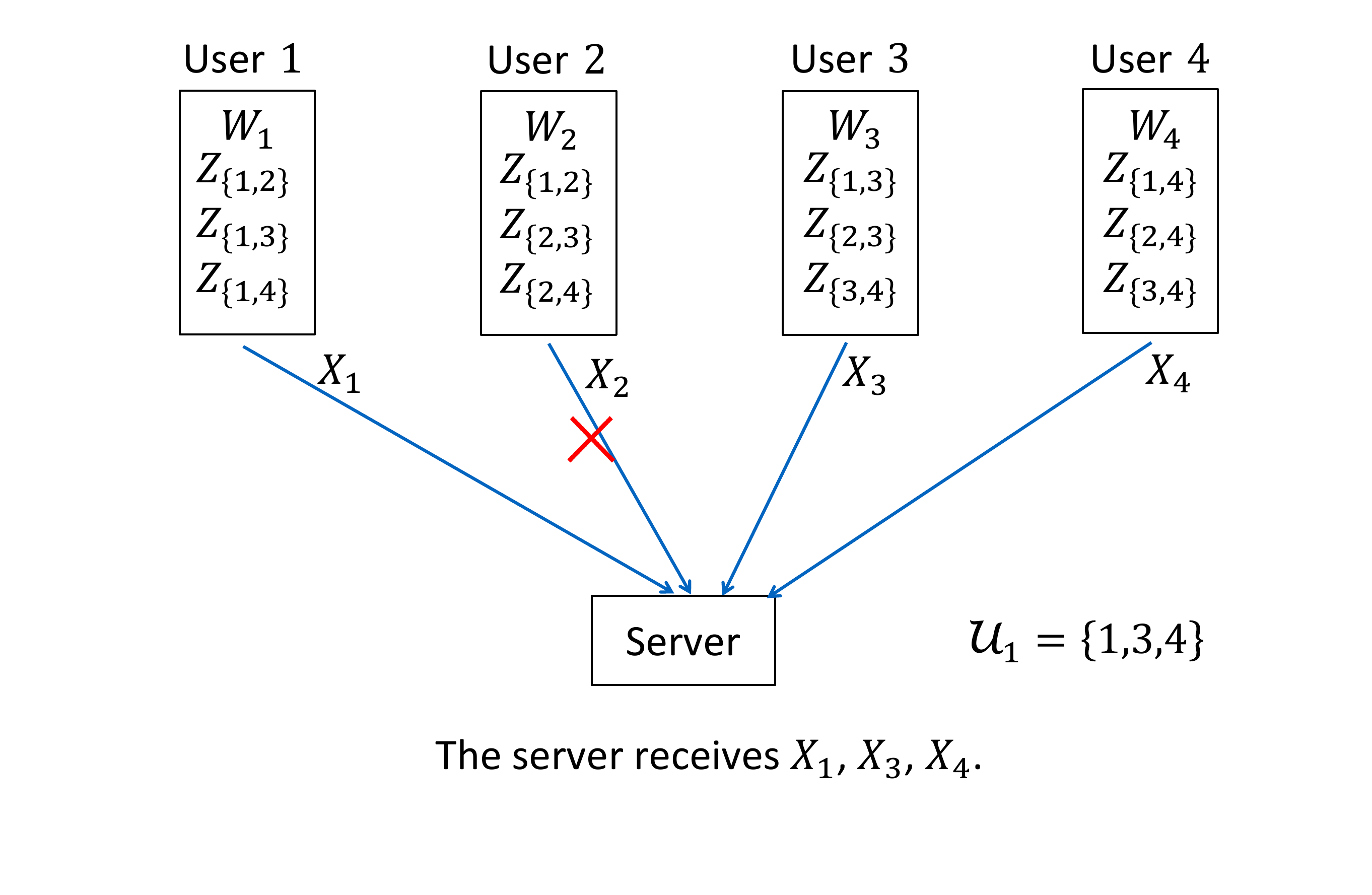}
        \caption{\small First round.}
        \label{fig:numerical 0a}
    \end{subfigure}%
   \\
    \begin{subfigure}[t]{1\textwidth}
        \centering
        \includegraphics[scale=0.25]{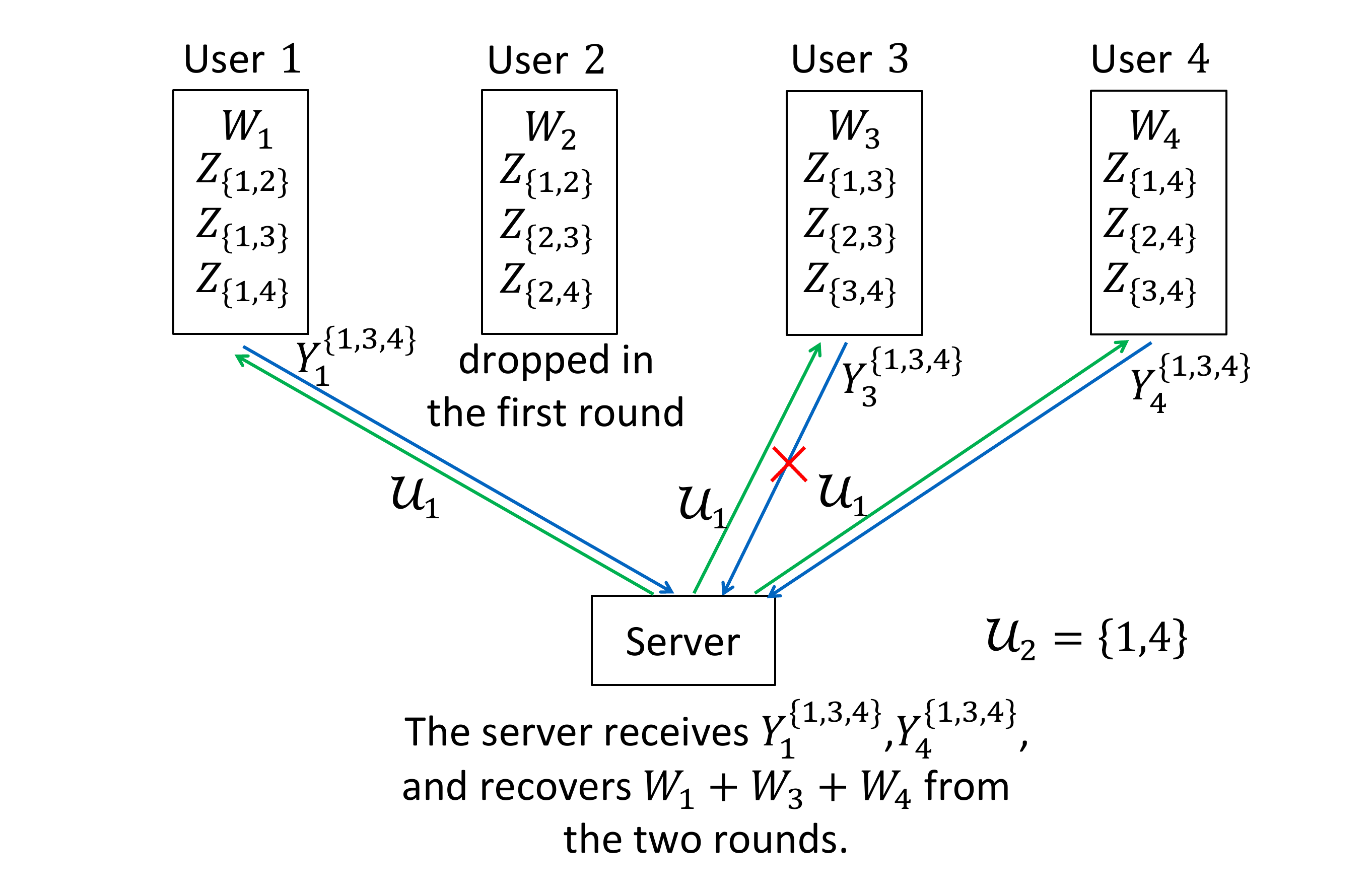}
        \caption{\small Second round.}
        \label{fig:numerical 0b}
    \end{subfigure}
    \caption{\small  $(\Ksf,\Usf,\Ssf)=(4,2,2)$ information theoretic secure aggregation problem with uncoded groupwise keys.}
    \label{fig:system model}
\end{figure}

Recently a modified version of the above problem, referred to as $(\Ksf,\Usf,\Ssf )$  information theoretic secure aggregation with uncoded groupwise keys, was proposed in~\cite{groupwisekey2022wan} as illusrated in Fig.~\ref{fig:system model}. An additional constraint on the keys was considered, where the key sharing among the users is ``uncoded'' and ``groupwise''. More precisely, 
given a system parameter $\Ssf$, the system generates ${\Ksf \choose \Ssf}$ mutually independent keys, 
such that each key is shared exactly by one group of $\Ssf$ distinct users and is also independent of the input vectors.\footnote{\label{foot:uncoded}In contrast, coded pairwise keys were used in~\cite{lightsec2021so,bonawitz2017practical} where the keys shared by the users are not mutually independent.}
 This constraint is motivated by the fact  that in practical cryptographic system, each key may be generated by some key agreement protocol as in~\cite{hellman1976newdirection,maurer1993secretkey,ahlswede1993commonran,csiszar2004secrey,gohari2010itkeyaggre,sun2020securegroupcast,sun2020compound} among a set of nodes.   
An interesting question arises for the $(\Ksf,\Usf,\Ssf )$  information theoretic secure aggregation with uncoded groupwise keys: do the capacity region   remains the same as the secure aggregation problem in~\cite{ITsecureaggre2021}?  
When $\Ssf> \Ksf-\Usf$, a secure aggregation scheme with groupwise keys was proposed in~\cite{groupwisekey2022wan} which achieves the same capacity region $\{(\Rsf_1,\Rsf_2):\Rsf_1 \geq 1,\Rsf_2 \geq 1/\Usf \}$  as  in~\cite{ITsecureaggre2021}.
Hence, in this case, the key group sharing constraint does not involve any loss of optimality.  
   When $\Ssf\leq \Ksf-\Usf$, a converse bound  was proposed in~\cite{groupwisekey2022wan} showing that the  capacity region   in~\cite{ITsecureaggre2021} is not achievable. However, the capacity region for the case $\Ssf\leq \Ksf-\Usf$ still remains open.

\subsection{Main Contribution}
The main contribution of this  paper is to  characterize the capacity region on the rate tuples for the $(\Ksf,\Usf,\Ssf )$  information theoretic secure aggregation with uncoded groupwise keys, $\left\{(\Rsf_1,\Rsf_2): \Rsf_1 \geq \frac{\binom{\Ksf-1}{\Ssf-1}}{\binom{\Ksf-1}{\Ssf-1}-\binom{\Ksf-1-\Usf}{\Ssf-1}}, \Rsf_2\geq \frac{1}{\Usf} \right\}$. More precisely, our focus is on the open case $\Ssf\leq \Ksf-\Usf$, and we develop the following results. 
\begin{itemize}
\item We first derive a new converse bound on the rate of the first   round $\Rsf_1$, which is strictly tighter than the converse bound in~\cite{groupwisekey2022wan}.  
\item We propose a new secure aggregation scheme based on the interference alignment strategy, which achieves the converse bound. The main difference between the proposed scheme and the secure aggregation scheme in~\cite{groupwisekey2022wan} (which works for the case $\Ssf>\Ksf-\Usf$) is that, when $\Ssf\leq \Ksf-\Usf$ the optimal rate in the first round $\Rsf_1$ is strictly larger than $1$, and  in addition to the masked input vector, each user should also transmit some additional messages composed of the keys. Hence, when a whole group of users sharing the same key drops in the second phase, the server can leverage these additional messages transmitted in the first round to decrypt.
\item We implement the proposed secure aggregation scheme into  the Tencent Cloud. Experimental results show that the proposed secure aggregation
scheme reduces the  model aggregation time by up to  $67.2\%$  compared
to the original secure aggregation scheme in~\cite{bonawitz2017practical}. 
\end{itemize}

\subsection{Paper Organization}
The rest of this paper is organized as follows. Section~\ref{sec:system} reviews the information theoretic  secure aggregation problem with uncoded groupwise keys. Section~\ref{sec:main} introduces the main theorem of the paper, which characterizes the capacity region.
Sections~\ref{sec:converse bound} and~\ref{sec:achievable bound} present the converse and achievability proofs of the main theorem, respectively. 
Section~\ref{sec:conclusion} concludes the paper, while some proofs are given in the Appendices.

\subsection{Notation Convention}
\label{sub:notation}
Calligraphic symbols denote sets, 
bold symbols denote vectors and matrices,
and sans-serif symbols denote system parameters.
We use $|\cdot|$ to represent the cardinality of a set or the length of a vector;
$[a:b]:=\left\{ a,a+1,\ldots,b\right\}$ and $[n] := [1:n]$; 
$\mathbb{F}_{\qsf}$ represents a  finite field with order $\qsf$;  
 $\ev_{n,i}$ represents the vertical $n$-dimensional unit vector  whose entry in the  $i^{\text{th}} $ position is $1$ and $0$ elsewhere; 
$\mathbf{A}^{\text{\rm T}}$  and $\mathbf{A}^{-1}$ represent the transpose  and the inverse of matrix $\mathbf{A}$, respectively;
$\text{rank}(\mathbf{A})$ represents the rank of matrix $\mathbf{A}$; 
$\mathbf{I}_n$ represents the identity matrix  of  dimension $n \times n$;
${\bf 0}_{m,n}$ represents all-zero matrix  of  dimension $m \times n$;
${\bf 1}_{m,n}$ represents all-one matrix  of  dimension $m \times n$;
$(\mathbf{A})_{m \times n}$ explicitly indicates that the matrix $\mathbf{A}$ is of dimension $m \times n$;
$<\cdot>_a$ represents the modulo operation with  integer quotient $a>0$ and in this paper we let $<\cdot>_a \in \{1,\ldots,a \}$ (i.e., we let $<b>_a=a$ if $a$ divides $b$);
  let $\binom{x}{y}=0$ if $x<0$ or $y<0$ or $x<y$;
let $\binom{\Xc}{y}=\{\Sc \subseteq \Xc: |\Sc|=y \}$ where $|\Xc|\geq y>0$.
In this paper, for a set  of real numbers  $\Sc$, we sort the elements in $\Sc$ in an increasing order and denote the $i^{\text{th}}$ smallest element by $\Sc(i)$, i.e., $\Sc(1)<\ldots<\Sc(|\Sc|)$.
  In the rest of the paper, entropies will be in base $\qsf$, where $\qsf$ represents the field size.

 \section{System Model}
\label{sec:system}
 We consider a $(\Ksf,\Usf, \Ssf)$ information theoretic secure aggregation problem with uncoded groupwise keys  originally formulated in~\cite{groupwisekey2022wan}, as illustrated in Fig~\ref{fig:system model}. 
Note that  $\Ksf,\Usf, \Ssf$ are given system parameters, where
  $\Ksf$ represents the number of users in the system, $\Usf$ represents the minimum number of surviving users, and $\Ssf$ represents the group-sharing parameter, i.e., the size of the groups uniquely sharing the same key. 
 Each user $k\in [\Ksf]$ holds one input vector $W_k$ containing $\Lsf$ uniform and i.i.d. symbols on a finite field $\mathbb{F}_{\qsf}$, where $\qsf$ is a prime power. 
In addition, for each set $\Vc\in \binom{[\Ksf]}{\Ssf}$, the users in $\Vc$ share a common key $Z_{\Vc}$ with large enough size. 
Considering that the key sharing is offline, the keys and the input vectors are assumed to be mutually independent, i.e., 
\begin{align}
H\left( \left(Z_{\Vc}:  \Vc \in \binom{[\Ksf]}{\Ssf}\right), (W_1,\ldots,W_{\Ksf})  \right) = \sum_{ \Vc \in \binom{[\Ksf]}{\Ssf} } H(Z_{\Vc}) +\sum_{k\in[\Ksf]} H(W_k). \label{eq:key constraint}
\end{align} 
We define  
$
Z_k := \left(Z_{\Vc}: \Vc \in \binom{[\Ksf]}{\Ssf}, k \in \Vc \right), 
$
as  the keys accessible by user $k\in [\Ksf]$.

A server is connected with the users via dedicated error-free links.\footnote{In this context, the nature of these links is irrelevant. Federated learning is a distributed process running at the  application layer, i.e., on top of some possibly heterogeneous networks with possibly different lower protocol layers. In any case, at the application layer, it is reasonable and practical to assume that the server and the users establish end-to-end communication sessions using TCP/IP. In fact, in most practical applications federated learning  runs over geographically widely separated users (imagine to run such application on the local picture library of smartphones all over the world).}  
 The server aims to aggregates the input vectors computed by the users. 
 In this paper, we consider the effect of user dropouts, i.e., the system is designed to tolerate up to $\Ksf - \Usf > 0$ user dropouts; in this case, it was proved in~\cite{ITsecureaggre2021} that one transmission round in the model aggregation is not enough. Thus we consider the two-round model aggregation as in~\cite{ITsecureaggre2021,groupwisekey2022wan}.

{\it First round.}
In the first round, each user $k\in [\Ksf]$ sends a coded message $X_k$ to the server  without knowing which user will drop in the future,  where  $X_k$ is completely determined by $W_k$ and $Z_k$, i.e, 
\begin{align}
H(X_k| W_k, Z_k)=0. \label{eq:HXk}
\end{align}
 The transmission rate of the first round $\Rsf_1$ is defined as the largest transmission load among all users normalized by $\Lsf$, i.e., 
\begin{align}
\Rsf_1:= \max_{k\in [\Ksf]} \frac{H\left(X_k\right)}{\Lsf}. \label{eq:def of R1}
\end{align}

Users may drop during the first round. We denote the set of surviving users after the first round by $\Uc_1$. Since $\Usf$ represents the minimum number of surviving users, we have    $\Uc_1 \subseteq [\Ksf]$ and $|\Uc_1|\geq \Usf$. Hence,  the server receives  $(X_k : k\in \Uc_1)$. 

{\it Second round.}
 In the second round, the server first sends the list of the surviving users  $\Uc_1$  to the users in $\Uc_1$. 
According to this information, each user $k\in \Uc_1$ sends another coded message    $Y^{\Uc_1}_{k}$ to the server, where 
\begin{align}
H(Y^{\Uc_1}_{k}| W_k, Z_k,\Uc_1)=0. \label{eq:Yk}
\end{align}
 The transmission rate of the second round $\Rsf_2$ is defined as the largest transmission load among all $\Uc_1$ and all users in $\Uc_1$ normalized by $\Lsf$, i.e.,
 \begin{align}
 \Rsf_2 := \max_{\Uc_1 \subseteq [\Ksf]:|\Uc_1|\geq \Usf} \ \max_{k\in \Uc_1}  \frac{H\left(Y^{\Uc_1}_{k}\right)}{\Lsf}. \label{eq:def of R2}
 \end{align}
 
Users may also drop during the second round transmission,  and the set of surviving users after the second round is denoted as $\Uc_2$. By definition, we have    $\Uc_2 \subseteq \Uc_1$ and $|\Uc_2|\geq \Usf$. Thus  the server receives  $Y^{\Uc_1}_{k}$  where $k\in \Uc_2$. 

 {\it Decoding.}
 From the two-round transmissions, the server totally receives $( X_{k_1} : k_1 \in \Uc_1 )$ and $( Y^{\Uc_1}_{k_2}: k_2 \in \Uc_2 )$, from which the server should recover  the sum of input vectors by the first round surviving users, i.e., $\sum_{k\in \Uc_1} W_k$. Thus 
\begin{align}
H\left( \sum_{k\in \Uc_1} W_k \Big|  ( X_{k_1} : k_1 \in \Uc_1 ), ( Y^{\Uc_1}_{k_2}: k_2 \in \Uc_2 ) \right)=0, \ \forall \Uc_1 \subseteq [\Ksf], \Uc_{2} \subseteq \Uc_1 : |\Uc_1| \geq  |\Uc_2|\geq \Usf. \label{eq:decodability}
\end{align}

 {\it Security.}
For the security constraint, we consider the worst-case, where the users are not really dropped but too slow in the transmission and thus the server may receive all the possible transmissions by the users. More precisely, 
it may receive  $(X_{k_1} : k_1 \in [\Ksf])$ from the first round and  $( Y^{\Uc_1}_{k_2}: k_2 \in \Uc_1 )$ from the second transmission. By security, from the received messages, the server can only obtain the computation task without retrieving other information about the input vectors, i.e., 
\begin{align}
I\left(   W_1,\ldots,W_{\Ksf} ; X_1,\ldots, X_{\Ksf}, ( Y^{\Uc_1}_{k}: k \in \Uc_1 )   \Big|  \sum_{k\in \Uc_1} W_k \right)=0 , \ \forall \Uc_1 \subseteq [\Ksf] : |\Uc_1|\geq \Usf. \label{eq:security constraint}
\end{align}
It is worth noticing that this mutual information is strictly equal to $0$; that is, we consider zero leakage, instead of vanishing leakage as $\Lsf\to \infty$.

{\it Objective.}
A rate tuple $(\Rsf_1,\Rsf_2)$ is achievable if there exist keys $\left(Z_{\Vc}:  \Vc \in \binom{[\Ksf]}{\Ssf} \right) $ satisfying~\eqref{eq:key constraint} and a secure aggregation scheme satisfying the decodability and security constraints in~\eqref{eq:decodability} and~\eqref{eq:security constraint}, respectively.  Our objective is to determine the capacity region (i.e., the closure of all achievable rate tuples), denoted by $\Rc^{\star}$.

{\it Existing converse bounds.}
By removing the uncoded groupwise constraint on the keys in our  problem, we obtain the information theoretic aggregation problem in~\cite{ITsecureaggre2021}. Hence, the converse bound  on the capacity region in~\cite{ITsecureaggre2021}  is also a converse bound for our  problem.
\begin{thm}[\cite{ITsecureaggre2021}]
\label{lem:converse}
For the $(\Ksf,\Usf, \Ssf)$ information theoretic secure aggregation problem with uncoded groupwise keys, any achievable rate tuple $(\Rsf_1,\Rsf_2)$ satisfies  
\begin{align}
\Rsf_1 \geq 1, \ \Rsf_2 \geq \frac{1}{\Usf}. \label{eq:converse from litterature} 
\end{align}
\hfill $\square$ 
\end{thm}

Considering the     uncoded groupwise constraint, an improved converse bound was given in~\cite{groupwisekey2022wan} for the case $\Ssf \leq  \Ksf-\Usf$. 
\begin{thm}[\cite{groupwisekey2022wan}]
\label{thm:subregion}
For the $(\Ksf,\Usf, \Ssf)$ information theoretic secure aggregation problem with uncoded groupwise keys, when $1=\Ssf \leq  \Ksf-\Usf$, secure aggregation is not possible; when  $2\leq \Ssf \leq  \Ksf-\Usf$,  any achievable rate tuple $(\Rsf_1,\Rsf_2)$ satisfies  
\begin{align}
\Rsf_1\geq  1+\frac{1}{  \binom{\Ksf-1}{\Ssf-1}-1}, \Rsf_2 \geq \frac{1}{\Usf} .\label{eq:converse of R1}
\end{align} 
\hfill $\square$ 
\end{thm}

{\it Existing achievable bound.}
A secure aggregation scheme with uncoded groupwise keys was proposed in~\cite{groupwisekey2022wan} for the case  $\Ssf > \Ksf-\Usf$, achieving the following rates. 
\begin{thm}[\cite{groupwisekey2022wan}]
\label{thm:ach result}
For the $(\Ksf,\Usf, \Ssf)$ information theoretic secure aggregation problem with uncoded groupwise keys, when $\Ssf > \Ksf-\Usf$, the following rate tuples are achievable, 
\begin{align}
\Rsf_1 \geq 1, \Rsf_2\geq \frac{1}{\Usf} . \label{eq:S>K-U result thm}
\end{align}
\hfill $\square$ 
\end{thm}
Comparing the achievable bound in Theorem~\ref{thm:ach result} and the converse bound in Theorem~\ref{lem:converse}, we can characterize the capacity region for the case $\Ssf > \Ksf-\Usf$, which is 
\begin{align}
\Rc^{\star} = \left\{(\Rsf_1,\Rsf_2): \Rsf_1 \geq 1, \Rsf_2\geq \frac{1}{\Usf} \right\}. \label{eq:S>K-U capacity}
\end{align}
However, no achievable scheme has been provided for the case $\Ssf\leq \Ksf - \Usf$, and the capacity region for this case remained open until this  paper.


 \section{Main Result}
\label{sec:main}
The main contribution of this paper is to fully characterize the capacity region for the information theoretic secure aggregation problem with uncoded groupwise keys. This is stated in the following theorem. 
\begin{thm}
\label{thm:main result}
For the $(\Ksf,\Usf, \Ssf)$ information theoretic secure aggregation problem with uncoded groupwise keys,  when $\Ssf=1$, secure aggregation is not possible; when  $\Ssf\geq 2$, we have 
\begin{align}
\Rc^{\star} = \left\{(\Rsf_1,\Rsf_2): \Rsf_1 \geq \frac{\binom{\Ksf-1}{\Ssf-1}}{\binom{\Ksf-1}{\Ssf-1}-\binom{\Ksf-1-\Usf}{\Ssf-1}}, \Rsf_2\geq \frac{1}{\Usf} \right\}. \label{eq:main result thm}
\end{align}
\hfill $\square$ 
\end{thm}
 The converse proof of Theorem~\ref{thm:main result} is given in Section~\ref{sec:converse bound} and the achievability proof is given in Section~\ref{sec:achievable bound}. 
The following remarks on Theorem~\ref{thm:main result} are in order:  
 \begin{itemize}
 \item  When $\Ssf> \Ksf-\Usf$, we have  $\binom{\Ksf-1-\Usf}{\Ssf-1}=0$ and thus the capacity region in~\eqref{eq:main result thm} reduces to the one in~\eqref{eq:S>K-U capacity}, which is also equal to the capacity region for the information theoretic secure aggregation problem in~\cite{ITsecureaggre2021} (the one without the constraint on the uncoded groupwise keys).   When $2\leq \Ssf\leq  \Ksf-\Usf$, the  additional communication rate from the optimal secure aggregation scheme with uncoded groupwise keys compared to the generally optimal secure aggregation scheme in~\cite{ITsecureaggre2021} is only at the first round  and is equal to 
 $
 \frac{\binom{\Ksf-1-\Usf}{\Ssf-1}}{\binom{\Ksf-1}{\Ssf-1}-\binom{\Ksf-1-\Usf}{\Ssf-1}}. 
 $
 \item The converse bound to prove~\eqref{eq:main result thm}  is strictly tighter than the existing one in~\eqref{eq:converse of R1}.  
 \item The secure aggregation scheme proposed in Section~\ref{sec:achievable bound} is a new and unified scheme working for all system parameters when $\Ssf>1$. In constrast, in~\cite{groupwisekey2022wan} only the regime $\Ssf>\Ksf-\Usf$ was  considered, where this regime was first divided into three cases and a different secure aggregation scheme was proposed for each case. 
 \item  The scheme in  Section~\ref{sec:achievable bound} uses all the $ \binom{\Ksf}{\Ssf}$ keys in the system , where each key contains $\frac{\Ssf\Lsf}{\binom{\Ksf-1}{\Ssf-1}-\binom{\Ksf-1-\Usf}{\Ssf-1}}$ symbols. In comparison,  the secure aggregation in~\cite{groupwisekey2022wan} uses  at most $\Oc(\Ksf^2)$   keys, where each   key has $\frac{(\Ksf-\Usf+1)\Lsf}{\Usf}$ symbols.
\end{itemize}

\section{Converse Proof of Theorem~\ref{thm:main result}}
\label{sec:converse bound}
Since the converse bound for the information theoretic aggregation problem in~\cite{ITsecureaggre2021} is also a converse bound for our considered problem,  by Theorem~\ref{lem:converse}  we can directly obtain  $\Rsf_2 \geq \frac{1}{\Usf}$. Hence, for Theorem~\ref{thm:main result} we only need to prove 
\begin{align}
\Rsf_1 \geq \frac{\binom{\Ksf-1}{\Ssf-1}}{\binom{\Ksf-1}{\Ssf-1}-\binom{\Ksf-1-\Usf}{\Ssf-1}}, \label{eq:converse bound of R1}
\end{align}
for the case $\Ssf \geq 2$. 

Let us focus on user $1$ and derive the lower bound of  $ \frac{H(X_1)}{\Lsf}$. 
For each set $\Uc\in \binom{[2:\Ksf]}{\Usf}$, letting $\Uc_1=\{1\} \cup \Uc$  and $\Uc_2=\Uc$, 
by the decodability constraint~\eqref{eq:decodability} the server can 
recover $\sum_{i\in \{1\} \cup \Uc} W_{i}$ from  $\left(X_{k}:k\in  \{1\} \cup \Uc\right)$ and $\left( Y^{\Uc_1}_{k}: k\in \Uc \right)$.
In addition,  for each $k\in \Uc$, $\left(X_{k},Y^{\Uc_1}_{k} \right)$ is a function of $(Z_{k}, W_{k})$.
Hence, the server can recover $\sum_{i\in \{1\} \cup \Uc} W_{i}$ from
  $(X_1,(Z_{k}, W_{k}:k\in \Uc ) )$; thus 
  \begin{subequations}
     \begin{align}
& 0=  H\left(\sum_{i\in \{1\} \cup \Uc} W_{i} \Bigg| X_1,(Z_{k}, W_{k}:k\in \Uc ) \right) \\
&= H\left( W_1 | X_1,(Z_{k}, W_{k}:k\in \Uc ) \right)\\
&=H\left(W_1 \Big| X_1, \Big(Z_{\Vc}: \Vc\in \binom{[\Ksf]}{\Ssf}, 1\in \Vc, \Uc\cap \Vc \neq \emptyset \Big)  \right) \label{eq:independent Zvs} \\
&= H\left(W_1 \Big| \Big(Z_{\Vc}: \Vc\in \binom{[\Ksf]}{\Ssf}, 1\in \Vc, \Uc\cap \Vc \neq \emptyset \Big)\right)- I\left(X_1;W_1 \Big| \Big(Z_{\Vc}: \Vc\in \binom{[\Ksf]}{\Ssf}, 1\in \Vc, \Uc\cap \Vc \neq \emptyset \Big)\right), 
 \label{eq:Hw1 given U}
   \end{align}
   \end{subequations}
   where~\eqref{eq:independent Zvs} comes from~\eqref{eq:key constraint} and~\eqref{eq:HXk}.
  In addition, from~\eqref{eq:Hw1 given U} we have 
  \begin{subequations}
  \begin{align}
&  I\left(X_1;W_1 \Big| \Big(Z_{\Vc}:\Vc\in \binom{[\Ksf]}{\Ssf}, 1\in \Vc, \Uc\cap \Vc \neq \emptyset \Big)\right)\nonumber\\
&= H\left(W_1 \Big| \Big(Z_{\Vc}: \Vc\in \binom{[\Ksf]}{\Ssf}, 1\in \Vc, \Uc\cap \Vc \neq \emptyset \Big)\right) \\
  &=H(W_1)=\Lsf, \label{eq:I=L}
  \end{align}
    \end{subequations}
    where~\eqref{eq:I=L} follows since $W_1$ is independent of the keys. 
   In addition, we have 
   \begin{subequations}
   \begin{align}
&   I\left(X_1; \Big(Z_{\Vc}: \Vc\in \binom{[\Ksf]}{\Ssf},1\in \Vc, \Uc\cap \Vc \neq \emptyset \Big) \Big| W_1 \right) 
  \nonumber\\& = I\left(X_1;  W_1,\Big(Z_{\Vc}: \Vc\in \binom{[\Ksf]}{\Ssf}, 1\in \Vc, \Uc\cap \Vc \neq \emptyset \Big)  \right) - I(X_1;W_1) \\
 &=   I\left(X_1;  W_1,\Big(Z_{\Vc}: \Vc\in \binom{[\Ksf]}{\Ssf}, 1\in \Vc, \Uc\cap \Vc \neq \emptyset \Big)  \right) \label{eq:IX1W1=0} \\
&\geq I\left(X_1; W_1 \Big| \Big(Z_{\Vc}:  \Vc\in \binom{[\Ksf]}{\Ssf},1\in \Vc, \Uc\cap \Vc \neq \emptyset \Big)  \right) \\& =\Lsf,
 \label{eq:I(X1;W1)}
   \end{align}
    \end{subequations}
where~\eqref{eq:IX1W1=0} comes from the security constraint~\eqref{eq:security constraint} and~\eqref{eq:I(X1;W1)}   comes from~\eqref{eq:I=L}. 
   
We sort the sets $\Vc\in \binom{[\Ksf]}{\Ssf}$ where $1\in \Vc$ in the lexicographic order, denoted by $\Sc_{1,1},\ldots,\Sc_{1,\binom{\Ksf-1}{\Ssf-1}}$.
Then by the chain rule of mutual information, we have 
\begin{subequations}
\begin{align}
& \Rsf_1 \Lsf \geq H(X_1) \\
&\geq I\left(X_1; Z_{\Sc_{1,1}},\ldots, Z_{\Sc_{1,\binom{\Ksf-1}{\Ssf-1} }} |W_1\right) \\   
&=\sum_{i\in \left[ \binom{\Ksf-1}{\Ssf-1}\right]} I(X_1; Z_{\Sc_{1,i}}|W_1, Z_{\Sc_{1,1}},\ldots,Z_{\Sc_{1,i-1}}) 
\label{eq:expand I} \\
&=\left(\binom{\Ksf-1}{\Usf}-\binom{\Ksf-\Ssf}{\Usf}\right) \sum_{i\in \left[ \binom{\Ksf-1}{\Ssf-1}\right]}\frac{1}{\binom{\Ksf-1}{\Usf}-\binom{\Ksf-\Ssf}{\Usf}} I(X_1; Z_{\Sc_{1,i}}|W_1, Z_{\Sc_{1,1}},\ldots,Z_{\Sc_{1,i-1}}) \\
&=\sum_{\Uc\in \binom{[2:\Ksf]}{\Usf} } \ \sum_{i\in \left[ \binom{\Ksf-1}{\Ssf-1}\right] : \Uc \cap \Sc_{1,i} \neq \emptyset} \frac{1}{\binom{\Ksf-1}{\Usf}-\binom{\Ksf-\Ssf}{\Usf}}  I(X_1; Z_{\Sc_{1,i}}|W_1, Z_{\Sc_{1,1}},\ldots,Z_{\Sc_{1,i-1}}), \label{eq:key sum in the key step converse}
\end{align}
 \end{subequations}
where~\eqref{eq:key sum in the key step converse} follows since for each   $i\in \left[ \binom{\Ksf-1}{\Ssf-1}\right]$, there are exactly $\binom{\Ksf-1}{\Usf}-\binom{\Ksf-\Ssf}{\Usf}$ sets in $\binom{[2:\Ksf]}{\Usf}$, each of  which  intersects $\Sc_{1,i}$.

From~\eqref{eq:key sum in the key step converse}, we have 
\begin{subequations}
\begin{align}
&\Rsf_1 \Lsf \geq  \frac{1}{\binom{\Ksf-1}{\Usf}-\binom{\Ksf-\Ssf}{\Usf}} \sum_{\Uc\in \binom{[2:\Ksf]}{\Usf}} \ \sum_{i\in \left[ \binom{\Ksf-1}{\Ssf-1}\right] : \Uc \cap \Sc_{1,i} \neq \emptyset} I(X_1; Z_{\Sc_{1,i}}|W_1, Z_{\Sc_{1,1}},\ldots,Z_{\Sc_{1,i-1}}) \\
& =\frac{1}{\binom{\Ksf-1}{\Usf}-\binom{\Ksf-\Ssf}{\Usf}}\sum_{\Uc\in \binom{[2:\Ksf]}{\Usf}} \ \sum_{i\in \left[ \binom{\Ksf-1}{\Ssf-1}\right] : \Uc \cap \Sc_1(i) \neq \emptyset} \nonumber\\& \Big(   I(X_1, (Z_{\Sc_{1,j}}: j\in [i-1], \Uc \cap \Sc_{1,j}= \emptyset); Z_{\Sc_{1,i}}|  W_1, (Z_{\Sc_{1,j}}: j\in [i-1], \Uc \cap \Sc_{1,j}\neq \emptyset) )  \nonumber\\& - I((Z_{\Sc_{1,j}}: j\in [i-1], \Uc \cap \Sc_{1,j}= \emptyset); Z_{\Sc_{1,i}}|W_1, (Z_{\Sc_{1,j}}: j\in [i-1], \Uc \cap \Sc_{1,j}\neq \emptyset) ) \Big) \\
&= \frac{1}{\binom{\Ksf-1}{\Usf}-\binom{\Ksf-\Ssf}{\Usf}}\sum_{\Uc\in \binom{[2:\Ksf]}{\Usf}} \ \sum_{i\in \left[ \binom{\Ksf-1}{\Ssf-1}\right] : \Uc \cap \Sc_{1,i} \neq \emptyset}  \nonumber\\&  I(X_1, (Z_{\Sc_{1,j}}: j\in [i-1], \Uc \cap \Sc_{1,j}= \emptyset); Z_{\Sc_{1,i}}|W_1, (Z_{\Sc_{1,j}}: j\in [i-1], \Uc \cap \Sc_{1,j}\neq \emptyset) )     \label{eq:independent keys and input} \\
&\geq \frac{1}{\binom{\Ksf-1}{\Usf}-\binom{\Ksf-\Ssf}{\Usf}}\sum_{\Uc\in \binom{[2:\Ksf]}{\Usf}} \ \sum_{i\in \left[ \binom{\Ksf-1}{\Ssf-1}\right] : \Uc \cap \Sc_{1,i} \neq \emptyset}  I(X_1; Z_{\Sc_{1,i}}|W_1, (Z_{\Sc_{1,j}}: j\in [i-1], \Uc \cap \Sc_{1,j}\neq \emptyset) )  \\
&= \frac{1}{\binom{\Ksf-1}{\Usf}-\binom{\Ksf-\Ssf}{\Usf}}\sum_{\Uc\in \binom{[2:\Ksf]}{\Usf}}  I\left(X_1; \Big(Z_{\Sc_{1,j}}: j\in\left[ \binom{\Ksf-1}{\Ssf-1}\right], \Uc \cap \Sc_{1,j}\neq \emptyset \Big) \Big|W_1  \right)  \label{eq:by chain rule MI} \\
&\geq \frac{1}{\binom{\Ksf-1}{\Usf}-\binom{\Ksf-\Ssf}{\Usf}}\sum_{\Uc\in \binom{[2:\Ksf]}{\Usf}} \Lsf \label{eq:taking L} \\
&=  \frac{\binom{\Ksf-1}{\Usf}}{\binom{\Ksf-1}{\Usf}-\binom{\Ksf-\Ssf}{\Usf}} \Lsf, \label{eq:most complicated step}
\end{align}
\end{subequations}
where~\eqref{eq:independent keys and input} follows since the uncoded keys and input vectors are independent,~\eqref{eq:by chain rule MI} comes from the chain rule of mutual information, and~\eqref{eq:taking L} comes from~\eqref{eq:I(X1;W1)}. 

Thus we have 
\begin{subequations}
\begin{align}
\Rsf_1 &\geq \frac{\binom{\Ksf-1}{\Usf}}{\binom{\Ksf-1}{\Usf}-\binom{\Ksf-\Ssf}{\Usf}} \\
&  =\frac{\binom{\Ksf-1}{\Ssf-1}}{\binom{\Ksf-1}{\Ssf-1}-\binom{\Ksf-1-\Usf}{\Ssf-1}}, \label{eq:final converse}
\end{align}
\end{subequations}
 where~\eqref{eq:final converse} follows since
\begin{subequations}
\begin{align}
&\binom{\Ksf-1}{\Usf} \binom{(\Ksf-1)-\Usf}{\Ssf-1}=\binom{\Ksf-1}{\Ssf-1}\binom{(\Ksf-1)-(\Ssf-1)}{\Usf} =\binom{ \Ksf -\Ssf }{\Usf} \binom{\Ksf-1}{\Ssf-1} \\
&\Longleftrightarrow   \binom{\Ksf-1}{\Usf} \left(\binom{\Ksf-1}{\Ssf-1}-\binom{\Ksf-1-\Usf}{\Ssf-1}\right)=\binom{\Ksf-1}{\Ssf-1} \left( \binom{\Ksf-1}{\Usf}-\binom{\Ksf-\Ssf}{\Usf} \right) .\label{eq:transform binomial}
\end{align}
 \end{subequations}  
  Hence, we proved~\eqref{eq:converse bound of R1} and thus  the converse bound of Theorem~\eqref{thm:main result}.

\section{Achievability Proof of Theorem~\ref{thm:main result}}
\label{sec:achievable bound} 
In this section, we describe the proposed secure aggregation scheme achieving the rate region in  Theorem~\ref{thm:main result}. We first illustrate the main ideas of the proposed scheme through the following example. 
 \begin{example}[$(\Ksf,\Usf, \Ssf))=(5,2,3)$]
\label{ex:case 1}
\rm
We consider the   $(\Ksf,\Usf, \Ssf)=(5,2,3)$ information theoretic secure aggregation problem with uncoded groupwise keys. Note that in this example,  for the ease of illustration, we assume that 
the field size $\qsf$ is a large enough prime; it will be proved later that our proposed scheme works for arbitrary field size. 

 By the converse bound derived in Section~\ref{sec:converse bound}, we have 
$
\Rsf_1\geq \frac{\binom{\Ksf-1}{\Ssf-1}}{\binom{\Ksf-1}{\Ssf-1}-\binom{\Ksf-1-\Usf}{\Ssf-1}} = \frac{6}{5}
$
and $\Rsf_2\geq \frac{1}{\Usf}=\frac{1}{2}$. 
 Inspired by the converse bound, we divide each input vector $W_i$ where $i\in [\Ksf]$ into $\binom{\Ksf-1}{\Ssf-1}-\binom{\Ksf-1-\Usf}{\Ssf-1}=5$ non-overlapping and equal-length pieces, $W_{i}=\{W_{i,1},\ldots,W_{i,5}\}$. 
For each set $\Vc\in \binom{[\Ksf]}{\Ssf}$, we generate a key $Z_{\Vc}$ containing $\frac{\Ssf\Lsf}{\binom{\Ksf-1}{\Ssf-1}-\binom{\Ksf-1-\Usf}{\Ssf-1}}=\frac{3\Lsf}{5}$ symbols uniformly i.i.d. over $\mathbb{F}_{\qsf}$; let $Z_{\Vc}$ be shared by the users in $\Vc$.  In addition, we further divide each key $Z_{\Vc}$ into $\Ssf=3$ sub-keys, $Z_{\Vc}=\{Z_{\Vc,k}:k\in \Vc\}$, where each sub-key $Z_{\Vc,k}$ has $\frac{\Lsf}{5}$ symbols. 

From the converse bound we see that in the first round each user $k\in [5]$ should send more than   $\Lsf$ symbols, 
 while input vector $W_k$ contains $\Lsf$ symbols.
Thus, unlike the secure aggregation scheme in~\cite{groupwisekey2022wan} which has $\Rsf_1=1$, in the first round besides the encrypted input vector, we also need to transmit some coded messages composed of keys, to cope with the fact that   some keys cannot be transmitted in the second round due to   user dropouts.
For each key $\Zc_{\Vc}$, we select a $6$-length vector $\av_{\Vc}=[a_{\Vc,1},\ldots,a_{\Vc,6}]^{\text{\rm T}}$ which will serve as the coefficient vector of its sub-keys during the first round. 
The selection of these coefficient vectors to guarantee the encodability, decodability and security, is the most important step in the proposed secure aggregation scheme. 
We  denote  the sets $\Vc\in \binom{[\Ksf]}{\Ssf}$ where   $k \in \Vc$  by  $\Sc_{k,1},\ldots,\Sc_{k,\binom{\Ksf-1}{\Ssf-1}}$; denote 
the sets  in $\binom{[\Ksf]\setminus\{k\}}{\Ssf}$   by  
$\overline{\Sc}_{k,1},\ldots,\overline{\Sc}_{k,\binom{\Ksf-1}{\Ssf}}$.
 For the security and encodability, it will be explained later that the selection   has the following two properties respectively:
\begin{align}
& \left[\av_{\Sc_{k,1}}, \ldots, \av_{\Sc_{k,\binom{\Ksf-1}{\Ssf-1}}} \right] \ \text{ has rank equal to } \binom{\Ksf-1}{\Ssf-1}=6, \ \forall k\in [\Ksf]; \label{eq:property example security}\\ 
& \left[\av_{\overline{\Sc}_{k,1}}, \ldots, \av_{\overline{\Sc}_{k,\binom{\Ksf-1}{\Ssf}}}  \right] \ \text{ has rank equal to } \binom{\Ksf-2}{\Ssf-1}=3, \ \forall k\in [\Ksf]. \label{eq:property example encodability}
\end{align}
In order to guarantee~\eqref{eq:property example security} and~\eqref{eq:property example encodability}, we select the coefficient vectors by the following two steps:
\begin{itemize}
\item We first select each vector  $\av_{\Vc}$ for each $\Vc \in \binom{[\Ksf]}{\Ssf}$ where $1\in \Vc$. More precisely, we choose each element in  $\av_{\Vc}$ uniformly i.i.d. over $\mathbb{F}_{\qsf}$.  
 In this example, we let 
\begin{align*}
& \av_{\{1,2,3\}}=[0, 1, 0, 0, 1, 1]^{\text{\rm T}}, \ \av_{\{1,2,4\}}=[1, 0, 1, 1, 1, 1]^{\text{\rm T}},\ \av_{\{1,2,5\}}=[0, 0, 0, 1, 0, 1]^{\text{\rm T}},\\
& \av_{\{1,3,4\}}=[0, 1, 1, 1, 0, 1]^{\text{\rm T}},\ \av_{\{1,3,5\}}=[1, 1, 0, 1, 0, 1]^{\text{\rm T}},\ \av_{\{1,4,5\}}=[1, 0, 0, 0, 0, 1]^{\text{\rm T}}. 
\end{align*}
\item Then we fix each of the remaining vectors  by a linear combination  of the selected vectors in the first step. More precisely, to fix $\av_{\{2,3,4\}}$, 
we let $\av_{\{2,3,4\}}$ be a linear combination  of  $\av_{\{1,3,4\}}$, $\av_{\{1,2,4\}}$, and $\av_{\{1,2,3\}}$, where the coefficients are either $+1$ or $-1$ and   alternated,
\begin{align}
\av_{\{2,3,4\}}= \av_{\{1,3,4\}}- \av_{\{1,2,4\}} +\av_{\{1,2,3\}}.\label{eq:ex1 av234}
\end{align}
Similarly, for each $\Vc \in \binom{[2:\Ksf]}{\Ssf}$,  we let $\av_{\Vc}$ be the following linear combination of $\av_{\Vc\setminus\{k\} \cup \{1\}}$ where $k\in \Vc$, (recall that $\Vc(i)$ represents the $i^{\text{th}}$ smallest element in $\Vc$) 
\begin{align}
\av_{\Vc}= \sum_{i\in [3]} (-1)^{i-1} \av_{\Vc\setminus \{\Vc(i)\} \cup \{1\}}.\label{eq:ex1 av}
\end{align}
\end{itemize}
The detailed section on the coefficient vectors is given in Table~\ref{tab:ex1}. It can be checked that this selection has the two properties in~\eqref{eq:property example security} and~\eqref{eq:property example encodability}. 
The first property could be directly checked. For the second property, we have  
\begin{align}
\av_{\{3,4,5\}}= \av_{\{2,4,5\}}-\av_{\{2,3,5\}} +\av_{\{2,3,4\}}; \label{eq:av 345 for user 1}
\end{align}
thus the rank of $[\av_{\{2,3,4\}},\av_{\{2,3,5\}},\av_{\{2,4,5\}},\av_{\{3,4,5\}}]$ is equal to the rank of $[\av_{\{2,3,4\}},\av_{\{2,3,5\}},\av_{\{2,4,5\}}]$ which is equal to $3$. In addition, since 
\begin{align}
\av_{\{3,4,5\}}= \av_{\{1,4,5\}}-\av_{\{1,3,5\}} +\av_{\{1,3,4\}};\label{eq:av 345 for user 2}
\end{align}
thus the rank of $[\av_{\{1,3,4\}},\av_{\{1,3,5\}},\av_{\{1,4,5\}},\av_{\{3,4,5\}}]$ is equal to the rank of $[\av_{\{1,3,4\}},\av_{\{1,3,5\}},\av_{\{1,4,5\}}]$ which is equal to $3$. Similarly, we can also check that 
the rank of  $[\av_{\{1,2,4\}},\av_{\{1,2,5\}},\av_{\{1,4,5\}},\av_{\{2,4,5\}}]$ is equal to the rank of $[\av_{\{1,2,4\}},\av_{\{1,2,5\}},\av_{\{1,4,5\}}]$ which is equal to $3$; the rank of  $[\av_{\{1,2,3\}},\av_{\{1,2,5\}},\av_{\{1,3,5\}},\av_{\{2,3,5\}}]$ is equal to the rank of $[\av_{\{1,2,3\}},\av_{\{1,2,5\}},\av_{\{1,3,5\}}]$ which is equal to $3$; the rank of  $[\av_{\{1,2,3\}},\av_{\{1,2,4\}},\av_{\{1,3,4\}},\av_{\{2,3,4\}}]$ is equal to the rank of $[\av_{\{1,2,3\}},\av_{\{1,2,4\}},\av_{\{1,3,4\}}]$ which is equal to $3$. Thus the property is satisfied. 
We will show later that this selection     guarantees the encodability, decodability and security.

\begin{table*}
 \centering
\protect\caption{Choice of $6$-dimensional vectors $\av_{\Vc}$ in the  $(\Ksf,\Usf, \Ssf))=(5,2,3)$ information theoretic secure aggregation problem. }\label{tab:ex1}
\begin{tabular}{|c|c|c|c|}
\hline 
$\av_{\Vc}$   & Value & $\av_{\Vc}$   & Value 
 \tabularnewline  
\hline 
\hline 
  $\av_{\{1,2,3\}} $  &  $[0, 1, 0, 0, 1, 1]^{\text{\rm T}}$   &   $\av_{\{1,4,5\}} $    & $[1, 0, 0, 0, 0, 1]^{\text{\rm T}}$ \tabularnewline  
\hline 
  $\av_{\{1,2,4\}} $  &  $[1, 0, 1, 1, 1, 1]^{\text{\rm T}}$    &   $\av_{\{2,3,4\}} $    & $ \av_{\{1,3,4\}}- \av_{\{1,2,4\}} +\av_{\{1,2,3\}}=[-1, 2, 0, 0, 0, 1]^{\text{\rm T}}$  \tabularnewline  
\hline 
  $\av_{\{1,2,5\}} $  &  $[0, 0, 0, 1, 0, 1]^{\text{\rm T}}$      &   $\av_{\{2,3,5\}} $    &   $\av_{\{1,3,5\}}- \av_{\{1,2,5\}} +\av_{\{1,2,3\}}=[1, 2, 0, 0, 1, 1]^{\text{\rm T}}$   \tabularnewline 
\hline 
  $\av_{\{1,3,4\}} $  &  $[0, 1, 1, 1, 0, 1]^{\text{\rm T}}$    &   $\av_{\{2,4,5\}} $   & $\av_{\{1,4,5\}}- \av_{\{1,2,5\}} +\av_{\{1,2,4\}}=[2, 0, 1, 0, 1, 1]^{\text{\rm T}}$ \tabularnewline  
\hline  
  $\av_{\{1,3,5\}} $    & $[1, 1, 0, 1, 0, 1]^{\text{\rm T}}$  &   $\av_{\{3,4,5\}} $    & $\av_{\{1,4,5\}}- \av_{\{1,3,5\}} +\av_{\{1,3,4\}}=[0, 0, 1, 0, 0, 1]^{\text{\rm T}}$ \tabularnewline  
\hline 
\end{tabular}
\end{table*}

After the selection of the above coefficient vectors,  the transmission in the first round  by each user $k\in [\Ksf]$ can be divided into two parts (as explained before):
\begin{itemize}
\item The first part contains $\binom{\Ksf-1}{\Ssf-1}-\binom{\Ksf-1-\Usf}{\Ssf-1}=5$ linear combinations of pieces and sub-keys, where each linear combination contains $\Lsf/5$ symbols. For each $j\in [5]$, let user $k$ transmit 
\begin{align}
X_{k,j} = W_{k,j}+ \sum_{\Vc\in \binom{[5]}{3}: k\in \Vc} a_{\Vc,j} Z_{\Vc,k}. \label{eq:ex1 Xkj 1}
\end{align}
 \item The second part contains $\binom{\Ksf-1-\Usf}{\Ssf-1}=1$ linear combination of sub-keys with $\Lsf/5$ symbols; let user $k$ transmit 
 \begin{align}
 X_{k,6}=\sum_{\Vc\in \binom{[5]}{3}: k\in \Vc} a_{\Vc,6} Z_{\Vc,k}.\label{eq:ex1 Xkj 2}
 \end{align}
\end{itemize} 
Hence, user $k$  transmits $X_k=(X_{k,1},\ldots, X_{k,6})$, totally $6\Lsf/5$ symbols in the first  round. Since the selection of the coefficient vectors has the property in~\eqref{eq:property example security},  the rank of the sub-keys in $X_k$ is equal to the dimension of $X_k$ and thus from $X_k$ the server cannot get any information about $W_k$ (see~\cite[Appendix~C]{groupwisekey2022wan} for the formal proof).

Now we consider the case  $\Uc_1=[5]$, i.e., no user drops in the first  round.  From the first round, the server can recover 
\begin{align}
\sum_{k_1\in [5]} X_{k_1,j}= \sum_{k_2\in [5]} W_{k_2,j} + \sum_{\Vc\in \binom{[5]}{3}}   a_{\Vc,j} \underbrace{\sum_{k_3\in \Vc} Z_{\Vc,k_3}}_{:=Z_{\Vc}^{[5]}} \label{eq:ex to be recovered}
\end{align}
for each $j\in [5]$, and  recover 
\begin{align}
\sum_{k_1\in [5]} X_{k_1,6}=    \sum_{\Vc\in \binom{[5]}{3}}  a_{\Vc,6} Z_{\Vc}^{[5]}  .
\end{align}
Hence, the server should further recover the second term on the RHS of~\eqref{eq:ex to be recovered},  $\sum_{\Vc\in \binom{[5]}{3}}   a_{\Vc,j} Z_{\Vc}^{[5]}$ for $j\in [5]$,   in the second round.

In the second round, to achieve $\Rsf_2=1/2$, we divide each $Z_{\Vc}^{[5]}$ where $\Vc \in \binom{[5]}{3}$ into $2$ non-overlapping and equal-length coded keys, $Z_{\Vc}^{[5]}=\left\{ Z_{\Vc,1}^{[5]} , Z_{\Vc,2}^{[5]} \right\}$, where each coded key contains $\frac{\Lsf}{10}$ symbols. 
Hence, we can write the 
recovery task of the second round in the matrix form 
\begin{align}
\left[ \begin{array}{c}
F_1\\
\vdots \\
 F_{12 }
\end{array} \right]= {\bf F}  \left[ \begin{array}{c}
Z_{\{1,2,3\}, 1}^{[5]}\\
Z_{\{1,2,4\}, 1}^{[5]}\\
\vdots \\
Z_{\{3,4,5\}, 1}^{[5]}\\
Z_{\{1,2,3\}, 2}^{[5]}\\
Z_{\{1,2,4\}, 2}^{[5]}\\
\vdots \\
Z_{\{3,4,5\}, 2}^{[5]}
\end{array} \right]
\label{eq:ex1 matrix form second round task}
\end{align}
where (recall that ${\bf 0}_{m \times n}$ represents the zero  matrix with dimension $m\times n$)
\begin{align}
 {\bf F}   = \left[\begin{array}{c:c}
 \av_{\{1,2,3\}}, \av_{\{1,2,4\}},\ldots, \av_{\{3,4,5\}}  & {\bf 0}_{6 \times 10}     \\ \hdashline
{\bf 0}_{6 \times 10}  &  \av_{\{1,2,3\}}, \av_{\{1,2,4\}},\ldots, \av_{\{3,4,5\}}   \\  
 \end{array}
\right]
\end{align}
Note that $F_6=\sum_{\Vc\in \binom{[5]}{3}}  a_{\Vc,6} Z_{\Vc,1}^{[5]}$ and $F_{12}=\sum_{\Vc\in \binom{[5]}{3}}  a_{\Vc,6} Z_{\Vc,2}^{[5]}$ have been already recovered by the server from the first  round. 

We focus on each user $k\in [5]$, who should transmit $\binom{\Ksf-1}{\Ssf-1}-\binom{\Ksf-1-\Usf}{\Ssf-1}= 5$ linear combinations of $F_1,\ldots,F_{12}$ in the second round; in the matrix form these $5$ linear combinations  are 
\begin{align}
{\bf S}_k \left[\begin{array}{c}
F_1\\
\vdots \\
 F_{12 }
\end{array} \right], 
\end{align}
where ${\bf S}_k$ is a matrix with dimension $5\times 	12$. Note that for the encodability, user $k$ can only compute the coded keys $Z_{\Vc,j}^{[5]}$ where $k\in \Vc$; thus in the transmitted linear combinations  the coefficients of the coded keys which user $k $ cannot compute should be equal to $0$. 

For user $1$, 
 the columns of  ${\bf S}_1 {\bf F}$ with indices in $[7:10] \cup [17:20]$ should be ${\bf 0}_{5 \times 1}$, since these columns correspond  to 
 $$Z_{\{2,3,4\},1}, Z_{\{2,3,5\},1}, Z_{\{2,4,5\},1} , Z_{\{3,4,5\},1}, Z_{\{2,3,4\},2}, Z_{\{2,3,5\},2}, Z_{\{2,4,5\},2} , Z_{\{3,4,5\},2},$$
which cannot be computed by user $1$.  
Assume that the column-wise sub-matrix of ${\bf F}$ including the columns with indices in $[7:10] \cup [17:20]$ is ${\bf F}_1$ with dimension $12 \times 8$, where 
\begin{align}
{\bf F}_1= 
\left[\begin{array}{c:c}
 \av_{\{2,3,4\}}, \av_{\{2,3,5\}},\av_{\{2,4,5\}} , \av_{\{3,4,5\}}  & {\bf 0}_{6 \times 4}     \\ \hdashline
{\bf 0}_{6 \times 4}  &   \av_{\{2,3,4\}}, \av_{\{2,3,5\}},\av_{\{2,4,5\}} , \av_{\{3,4,5\}}   \\  
 \end{array}
\right].
\label{eq:F1 form}
\end{align}
We need to find   $5$ linearly independent  left null vectors of ${\bf F}_1$, and let ${\bf S}_1$ be the matrix of these $5$ vectors. Note that if ${\bf F}_1$ is full rank, the left null space of ${\bf F}_1$ only contains $12-8=4$ linearly independent vectors. However, by our construction, it has been shown in~\eqref{eq:av 345 for user 1} that  $\av_{\{3,4,5\}}= \av_{\{2,4,5\}}-\av_{\{2,3,5\}} +\av_{\{2,3,4\}}$; in other words, 
{\bf the coefficient vectors corresponding to the unknown coded keys of user $1$ are aligned.}
 Thus by this interference alignment-like construction leading to~\eqref{eq:property example encodability},  the rank of ${\bf F}_1$ is $6$, and thus the left null space of ${\bf F}_1$ contains $12-6=6$ linearly independent vectors. 
More precisely,   the left null space of $[ \av_{\{2,3,4\}}, \av_{\{2,3,5\}},\av_{\{2,4,5\}}]$ is the linear space spanned by 
$$\sv_{1,1}=(0, -1, -2, 0, 0, 2), \ \sv_{1,2}= (-2, -1, 0, 0, 4, 0), \ \sv_{1,3}= (0, 0, 0, 1, 0, 0).$$ Hence, the left null space of ${\bf F}_1$ is the linear space spanned by 
\begin{align}
(\sv_{1,1},{\bf 0}_{1\times 6}), \ (\sv_{1,2},{\bf 0}_{1\times 6}), \ (\sv_{1,3},{\bf 0}_{1\times 6}), \ ({\bf 0}_{1\times 6},\sv_{1,1}), \ ({\bf 0}_{1\times 6},\sv_{1,2}), \ ({\bf 0}_{1\times 6},\sv_{1,3}).
\label{eq:ex1 6 vectors}
\end{align}
We let each row of ${\bf S}_1$  be  one  random linear combination of the vectors in~\eqref{eq:ex1 6 vectors}; in this example, we let 
\begin{align}
{\bf S}_1= 
\left[ \begin{array}{cccccccccccc}
0& -1& -2& 0& 0& 2& 0& 0& 0& 0& 0& 0 \\
-2& -1& 0& 0& 4& 0& 0& 0& 0& 0& 0& 0\\
0& 0& 0& 0& 0& 0& 0& -1& -2& 0& 0& 2\\
0& 0& 0& 0& 0& 0& -2& -1& 0& 0& 4& 0\\
0& 0& 0& 1& 0& 0& 0& 0& 0& 1& 0& 0
\end{array} \right].
\label{eq:ex1 S1}
\end{align}

For user $2$,   the columns of  ${\bf S}_2 {\bf F}$ with indices in $\{4,5,6,10,14,15,16,20\}$ should be ${\bf 0}_{5 \times 1}$, since these columns correspond  to 
 $$Z_{\{1,3,4\},1}, Z_{\{1,3,5\},1}, Z_{\{1,4,5\},1} , Z_{\{3,4,5\},1}, Z_{\{1,3,4\},2}, Z_{\{1,3,5\},2}, Z_{\{1,4,5\},2} , Z_{\{3,4,5\},2},$$
which cannot be computed by user $2$.  
Assume that the column-wise sub-matrix of ${\bf F}$ including the columns with indices in $\{4,5,6,10,14,15,16,20\}$ is ${\bf F}_2$ with dimension $12 \times 8$, where 
\begin{align*}
{\bf F}_2= 
\left[\begin{array}{c:c}
 \av_{\{1,3,4\}}, \av_{\{1,3,5\}},\av_{\{1,4,5\}} , \av_{\{3,4,5\}}  & {\bf 0}_{6 \times 4}     \\ \hdashline
{\bf 0}_{6 \times 4}  &   \av_{\{1,3,4\}}, \av_{\{1,3,5\}},\av_{\{1,4,5\}} , \av_{\{3,4,5\}}   \\  
 \end{array}
\right].
\end{align*}
By construction we have $\av_{\{3,4,5\}}= \av_{\{1,4,5\}}-\av_{\{1,3,5\}} +\av_{\{1,3,4\}}$  as shown in~\eqref{eq:av 345 for user 2}. The left null space of $[ \av_{\{2,3,4\}}, \av_{\{2,3,5\}},\av_{\{2,4,5\}}]$ is the linear space spanned by 
$$\sv_{2,1}=(-1, 0, -1, 0, 0, 1), \ \sv_{2,2}= (0, 0, 0, 0, 1, 0), \ \sv_{2,3}= (0, -1, 0, 1, 0, 0).$$ Hence, the left null space of ${\bf F}_2$ is the linear space spanned by 
\begin{align*}
(\sv_{2,1},{\bf 0}_{1\times 6}), \ (\sv_{2,2},{\bf 0}_{1\times 6}), \ (\sv_{2,3},{\bf 0}_{1\times 6}), \ ({\bf 0}_{1\times 6},\sv_{2,1}), \ ({\bf 0}_{1\times 6},\sv_{2,2}), \ ({\bf 0}_{1\times 6},\sv_{2,3}).
\end{align*}
 We let each row of ${\bf S}_2$  be  one  random linear combination of the vectors in~\eqref{eq:ex1 6 vectors}; in this example, we let 
\begin{align}
{\bf S}_2= 
\left[ \begin{array}{cccccccccccc}
-1& 0& -1& 0& 0& 1& 0& 0& 0& 0& 0& 0\\
0& 0& 0& 0& 1& 0& 0& 0& 0& 0& 0& 0\\
0& 0& 0& 0& 0& 0& -1& 0& -1& 0& 0& 1\\
0& 0& 0& 0& 0& 0& 0& 0& 0& 0& 1& 0\\
0& -1& 0& 1& 0& 0& 0& -2& 0& 2& 0& 0
\end{array} \right].
\end{align}
Similarly, we let 
\begin{align}
&{\bf S}_3= 
\left[ \begin{array}{cccccccccccc}
-1& 0& 1& -1& 0& 1& 0& 0& 0& 0& 0& 0\\
0& 0& -1& 0& 1& 0& 0& 0& 0& 0& 0& 0\\
0& 0& 0& 0& 0& 0& -1& 0& 1& -1& 0& 1\\
0& 0& 0& 0& 0& 0& 0& 0& -1& 0& 1& 0\\
0& 1& 0& 0& 0& 0& 0& 3& 0& 0& 0& 0
\end{array} \right],\\
&{\bf S}_4= 
\left[ \begin{array}{cccccccccccc}
1& -1& 0& -1& 0& 1& 0& 0& 0& 0& 0& 0\\
1& -1& 0& 0& 1& 0& 0& 0& 0& 0& 0& 0\\
0& 0& 0& 0& 0& 0& 1& -1& 0& -1& 0& 1\\
0& 0& 0& 0& 0& 0& 1& -1& 0& 0& 1& 0\\
0& 0& 1& 0& 0& 0& 0& 0& 4& 0& 0& 0
\end{array} \right], \label{eq:ex1 S4} \\
&
{\bf S}_5= 
\left[ \begin{array}{cccccccccccc}
-1& -1& 0& 0& 0& 1& 0& 0& 0& 0& 0& 0\\
-2& -1& 1& 0& 1& 0& 0& 0& 0& 0& 0& 0\\
0& 0& 0& 0& 0& 0& -1& -1& 0& 0& 0& 1\\
0& 0& 0& 0& 0& 0& -2& -1& 1& 0& 1& 0\\
0& 0& -1& 1& 0& 0& 0& 0& -5& 5& 0& 0
\end{array} \right].
\label{eq:ex1 S5}
\end{align}
As a summary, the constraint~\eqref{eq:property example encodability} is satisfied by the   interference alignment-like construction, while satisfying this constraint leads to the successful encoding of each user. 

Then we check the decodability. Note  that $F_6$ and $F_{12}$ have been recovered by the server from the first  round. Recall that $\ev_{n,i}$ represents the vertical $n$-dimensional standard unit vector  whose    $i^{\text{th}} $ element is $1$.
 For any set of two users $\Uc_2=\{u_1,u_2\} \subseteq [\Ksf]$ where $|\Uc_2|=2$, one can check that  from~\eqref{eq:ex1 S1}-\eqref{eq:ex1 S5} that the matrix
 \begin{align}
 \left[\begin{array}{c}
{\bf S}_{u_1} \\
{\bf S}_{u_2} \\
\ev_{12,6}^{\text{\rm T}}\\
\ev_{12,12}^{\text{\rm T}}
 \end{array}
\right]
\label{eq:Su1 u2}
 \end{align}
whose dimension is $12\times 12$,  is full rank; thus the server can recover $F_1,\ldots,F_{12}$ and then recover $W_1+\cdots+W_{5}$. 

For the security, from the first  round the server cannot obtain any information about $W_1,\ldots, W_{5}$. In the second round,  all the transmissions by all users are linear combinations of $F_1,\ldots, F_{12}$, where $F_6$ and $F_{12}$ can be recovered from the first round. Since each $F_i$, where $i\in [12]\setminus \{6,12\}$ contains $\Lsf/10$ symbols, by~\cite{shannonsecurity} the server can only obtain additional $10 \Lsf/10 =\Lsf$ symbols about $W_1,\ldots, W_{5}$ from the second round, which are exactly the symbols in $W_1+\cdots+W_{5}$. Hence, the proposed secure aggregation scheme is secure. 
 
The above scheme could be directly extended to other $\Uc_1 \subseteq [5]$ where $\Uc_1 \geq 2$. For example, consider $\Uc_1=[4]$.  
After the first  round, the server can recover 
 \begin{align}
\sum_{k_1\in [4]} X_{k_1,j}= \sum_{k_2\in [4]} W_{k_2,j} + \sum_{\Vc\in \binom{[5]}{3}}   a_{\Vc,j} \underbrace{\sum_{k_3\in \Vc \cap [4]} Z_{\Vc,k_3}}_{:=Z_{\Vc}^{[4]}}
\end{align}
for each $j\in [5]$, and  recover 
\begin{align}
\sum_{k_1\in [4]} X_{k_1,6}=    \sum_{\Vc\in \binom{[5]}{3}}  a_{\Vc,6} Z_{\Vc}^{[4]}  .
\end{align}
Hence, the server should further recover in the second round 
 \begin{align}
\left[ \begin{array}{c}
F_1\\
\vdots \\
 F_{12 }
\end{array} \right]= {\bf F}  \left[ \begin{array}{c}
Z_{\{1,2,3\}, 1}^{[4]}\\
Z_{\{1,2,4\}, 1}^{[4]}\\
\vdots \\
Z_{\{3,4,5\}, 1}^{[4]}\\
Z_{\{1,2,3\}, 2}^{[4]}\\
Z_{\{1,2,4\}, 2}^{[4]}\\
\vdots \\
Z_{\{3,4,5\}, 2}^{[4]}
\end{array} \right],
\label{eq:ex1 matrix form second round task UC1 4}
\end{align}
where ${\bf F}$ is given in~\eqref{eq:F1 form}, and  $F_6 , F_{12}$ have been recovered from the first  round. By choosing ${\bf S}_1, {\bf S}_2, {\bf S}_3, {\bf S}_4$ as in~\eqref{eq:ex1 S1}-\eqref{eq:ex1 S4}, we let each user $k\in [4]$ transmit ${\bf S}_{k} {\bf F}$ in the second round. For any set of two users $\Uc_2=\{u_1,u_2\} \subseteq [4]$ where $|\Uc_2|=2$, the matrix in~\eqref{eq:Su1 u2} is full rank; thus the decodability was proved. 
By the same reason as the case $\Uc_1=[5]$, we can also prove that the proposed scheme is secure for the case $\Uc_1=[4]$.

As a result, the proposed secure aggregation scheme achieves $\Rsf_1=6/5$ and $\Rsf_2=1/2$, coinciding with the converse bound in Section~\ref{sec:converse bound}.
\hfill $\square$
\end{example}

We   are now ready to generalize Example~\ref{ex:case 1} and 
describe the two-round secure aggregation scheme for the case $\Ssf\leq \Ksf-\Usf$.
 Note that to design achievable schemes, 
as explained in~\cite{ITsecureaggre2021},   we can assume    that $\qsf$ is large enough  without loss of generality.


{\it First round.}
To achieve $\Rsf_1=\frac{\binom{\Ksf-1}{\Ssf-1}}{\binom{\Ksf-1}{\Ssf-1}-\binom{\Ksf-1-\Usf}{\Ssf-1}}$  coinciding with the  converse bound in Section~\ref{sec:converse bound}, in the first round  we divide each input vector $W_i$ where $i\in [\Ksf]$ into $ \binom{\Ksf-1}{\Ssf-1}-\binom{\Ksf-1-\Usf}{\Ssf-1} $ non-overlapping and equal-length pieces, $W_i=\left\{W_{i,j}:j\in \left[ \binom{\Ksf-1}{\Ssf-1}-\binom{\Ksf-1-\Usf}{\Ssf-1}  \right] \right\}$. Hence, each piece contains $\frac{\Lsf}{\binom{\Ksf-1}{\Ssf-1}-\binom{\Ksf-1-\Usf}{\Ssf-1}}$ symbols. 
In addition, for each  $\Vc \in \binom{[\Ksf]}{\Ssf}$ we   generate a key $Z_{\Vc}$ with $\frac{\Ssf \Lsf}{\binom{\Ksf-1}{\Ssf-1}-\binom{\Ksf-1-\Usf}{\Ssf-1}}$ symbols uniformly i.i.d. over $\mathbb{F}_{\qsf}$, shared by the users in $\Vc$. We further divide each key $Z_{\Vc}$ into $\Ssf$ non-overlapping and equal-length sub-keys, $Z_{\Vc}=\{Z_{\Vc,k}:k\in \Vc\}$, where each sub-key $Z_{\Vc,k}$ contains $\frac{\Lsf}{\binom{\Ksf-1}{\Ssf-1}-\binom{\Ksf-1-\Usf}{\Ssf-1}}$ symbols uniformly i.i.d. over $\mathbb{F}_{\qsf}$.


Note that, by the converse bound of Theorem~\eqref{thm:main result}, 
each user  $k\in [\Ksf]$  transmits     $\binom{\Ksf-1}{\Ssf-1}$ linear combinations of pieces and keys in the first round, while $W_k$ only contains $ \binom{\Ksf-1}{\Ssf-1}-\binom{\Ksf-1-\Usf}{\Ssf-1} $ pieces.\footnote{\label{foot:main difference}This is  the main step to deal with the chanllenge arised in the case $\Ssf\leq \Ksf-\Usf$ compared to the secure aggregation scheme in~\cite{groupwisekey2022wan} for the case $\Ssf>\Ksf-\Usf$. More precisely,  when $\Ssf\leq \Ksf-\Usf$, there may exist some key  for which the whole group of users knowing 
that key   all drop  after the second round.  Thus we need to transmit more than one (normalized) linear combinations in the first round to deal with this event, such that even if all the knowing users drop during the transmissions, we can still remove ``interference'' of  this key to recover the task.}
 Hence, in the first round of the proposed secure aggregation scheme, the  first-round transmission by user $k$, $X_k=\left\{X_{k,j}: j\in  \left[\binom{\Ksf-1}{\Ssf-1} \right] \right\}$, contains two parts:
\begin{itemize}
\item for $j\in \left[ \binom{\Ksf-1}{\Ssf-1}-\binom{\Ksf-1-\Usf}{\Ssf-1} \right]$, we construct 
\begin{align}
X_{k,j}=W_{k,j}+ \sum_{\Vc\in \binom{[\Ksf]}{\Ssf}: k\in \Vc} a_{\Vc,j} Z_{\Vc,k}; \label{eq:first part first round}
\end{align} 
\item for $j\in \left[ \binom{\Ksf-1}{\Ssf-1}-\binom{\Ksf-1-\Usf}{\Ssf-1}+1 : \binom{\Ksf-1}{\Ssf-1} \right]$, we construct 
\begin{align}
X_{k,j}=\sum_{\Vc\in \binom{[\Ksf]}{\Ssf}: k\in \Vc} a_{\Vc,j} Z_{\Vc,k}.  \label{eq:second part first round}
\end{align}
\end{itemize}
For each $\Vc \in \binom{[\Ksf]}{\Ssf}$, define $\av_{\Vc}=\left[a_{\Vc,1},a_{\Vc,2},\ldots,a_{\Vc,\binom{\Ksf-1}{\Ssf-1}}   \right]^{\text{\rm T}}$, where each $a_{\Vc,j}\in \mathbb{F}_{\qsf}$ is a coefficient to be clarified later. By the security constraint, from $X_k$, the server cannot get any information about $W_k$; thus 
by denoting  the sets $\Vc\in \binom{[\Ksf]}{\Ssf}$ where   $k \in \Vc$  by  $\Sc_{k,1},\ldots,\Sc_{k,\binom{\Ksf-1}{\Ssf-1}}$, we   aim to have that    the coefficients matrix  (whose dimension is   $\binom{\Ksf-1}{\Ssf-1}\times \binom{\Ksf-1}{\Ssf-1}$)
\begin{align}
\left[\av_{\Sc_{k,1}},\ldots, \av_{\Sc_{k,\binom{\Ksf-1}{\Ssf-1}}} \right] \ \text{ has rank equal to $\binom{\Ksf-1}{\Ssf-1}$}, \ \forall k\in[\Ksf]. \label{eq:full rank constraint}
\end{align}

The choice of $\av_{\Vc}$ where $\Vc \in \binom{[\Ksf]}{\Ssf}$ is the most non-trivial design in the proposed secure aggregation scheme, which should satisfy the security constraint~\eqref{eq:full rank constraint} and guarantee the
 encodability and decodability of the second-round transmission.  Our selection contains two steps:
 \begin{itemize}
 \item First step. For each $\Vc \in \binom{[\Ksf]}{\Ssf}$ where $1\in \Vc$, we let $\av_{\Vc}$ be uniform and i.i.d. over $\mathbb{F}^{\binom{\Ksf-1}{\Ssf-1}}_{\qsf}$. Thus we have fixed 
  $\binom{\Ksf-1}{\Ssf-1}$ coefficient vectors in random. 
 \item Second step. For each $\Vc \in \binom{[2:\Ksf]}{\Ssf}$, we let $\av_{\Vc}$ be a linear combination of some vectors  fixed in the first step. More precisely, we let 
 \begin{align}
 \av_{\Vc}= \sum_{i\in [\Ssf]} (-1)^{i-1}\av_{\Vc\setminus \{\Vc(i)\} \cup \{1\}}. \label{eq:construction of other av}
 \end{align}
 \end{itemize}

 The above choice leads to the following lemma, which is crucial for the second-round transmissions based on interference alignment.  
 \begin{lem}
 \label{lem:crucial lemma on transform}
 By the above choice of coefficient vectors, for each $\Vc \in \binom{[\Ksf]}{\Ssf}$ and each $k\in [\Ksf]\setminus \Vc$, denoting the number of elements in $\Vc$ smaller than $k$ by $n_{\Vc,k}$, we have 
 \begin{align}
 \av_{\Vc}= \sum_{i_1\in [n_{\Vc,k}+1:\Ssf]}(-1)^{i_1-n_{\Vc,k}-1}  \av_{\Vc\setminus \{\Vc(i_1)\} \cup \{k\}} + \sum_{i_2\in [n_{\Vc,k}] } (-1)^{n_{\Vc,k}+i_2} \av_{\Vc\setminus \{\Vc(i_2)\} \cup \{k\}}.\label{eq:general tranform eq}
\end{align} 
\hfill $\square$  
 \end{lem}
The proof of Lemma~\ref{lem:crucial lemma on transform} could be found in Appendix~\ref{sec:proof of crucial lemma on transform}. By Lemma~\ref{lem:crucial lemma on transform}, it can be directly seen that 
for each $k\in [\Ksf]$,  from the coefficient vectors $\av_{\Vc}$ where  $\Vc \in \binom{[\Ksf]}{\Ssf}$ and   $k\in \Vc $, we  can re-construct all $\binom{\Ksf}{\Ssf}$ coefficient vectors; thus the constraint~\eqref{eq:full rank constraint} is satisfied with high probability.\footnote{\label{foot:containging 1 full rank}We generate $\binom{\Ksf-1}{\Ssf-1}$ coefficient vectors $\av_{\Vc}$ where $\Vc \in \binom{[\Ksf]}{\Ssf}$ and $1\in \Vc$, and each vector contains $\binom{\Ksf-1}{\Ssf-1}$  elements  uniformly i.i.d. over a large enough finite field. Hence, these $\binom{\Ksf-1}{\Ssf-1}$ coefficient vectors are linearly independent with high probability.}

Due to the user dropouts,  the server only  receives $\{X_{k}:k\in \Uc_1\}$. Hence, for each $j\in \left[ \binom{\Ksf-1}{\Ssf-1}-\binom{\Ksf-1-\Usf}{\Ssf-1} \right]$, the server recovers 
\begin{align}
\sum_{k_1\in\Uc_1} X_{k_1,j}&=  \sum_{k_2\in\Uc_1} W_{k_2,j} + \sum_{\Vc\in \binom{[\Ksf]}{\Ssf}: \Vc\cap \Uc_1 \neq \emptyset}   a_{\Vc,j} \underbrace{\sum_{k_3\in \Vc\cap \Uc_1} Z_{\Vc,k_3}}_{:=Z_{\Vc}^{\Uc_1}} ;  
 \label{eq:first round sum first part} 
\end{align}
for each  $j\in \left[ \binom{\Ksf-1}{\Ssf-1}-\binom{\Ksf-1-\Usf}{\Ssf-1}+1 : \binom{\Ksf-1}{\Ssf-1} \right]$,  the server recovers 
\begin{align}
\sum_{k_1\in\Uc_1} X_{k_1,j}=    \sum_{\Vc\in \binom{[\Ksf]}{\Ssf}: \Vc\cap \Uc_1 \neq \emptyset}  a_{\Vc,j} Z_{\Vc}^{\Uc_1}  . \label{eq:round 2 knows} 
\end{align}

{\it Second round.}
The task of the  second round transmission is to let  each user $k\in \Uc_1$ transmit  $Y^{\Uc_1}_{k}$ such that from any subset of users $\Uc_2\subseteq \Uc_1$ where $|\Uc_1| \geq |\Uc_2|\geq  \Usf$, 
 the server can recover 
\begin{align}
  \sum_{\Vc\in \binom{[\Ksf]}{\Ssf}: \Vc\cap \Uc_1 \neq \emptyset}   a_{\Vc,j_1} Z_{\Vc}^{\Uc_1} , \ \forall j_1 \in \left[ \binom{\Ksf-1}{\Ssf-1}-\binom{\Ksf-1-\Usf}{\Ssf-1}\right]  ,\label{eq:round 2 task}
\end{align} 
  from $\{Y^{\Uc_1}_{k_1}:k_1 \in \Uc_2\}$ and $\left\{\sum_{k_1\in\Uc_1} X_{k_1,j} : j\in \left[ \binom{\Ksf-1}{\Ssf-1}-\binom{\Ksf-1-\Usf}{\Ssf-1}+1 : \binom{\Ksf-1}{\Ssf-1} \right] \right\}$ in~\eqref{eq:round 2 knows}.
 In other words, from $\{Y^{\Uc_1}_{k_1}:k_1 \in \Uc_2\}$ and the linear combinations in~\eqref{eq:round 2 knows}, the server should recover 
 \begin{align}
 \sum_{\Vc\in \binom{[\Ksf]}{\Ssf}: \Vc\cap \Uc_1 \neq \emptyset}   a_{\Vc,j} Z_{\Vc}^{\Uc_1} , \ \forall j  \in \left[ \binom{\Ksf-1}{\Ssf-1} \right]  .\label{eq:round 2 total task}
 \end{align}

To achieve $\Rsf_2=1/\Usf$, we further divide each  $Z_{\Vc}^{\Uc_1}$ where $\Vc \in \binom{[\Ksf]}{\Ssf}$ and  $\Vc\cap \Uc_1 \neq \emptyset$, into $\Usf$ non-overlapping and equal-length coded keys, 
$Z_{\Vc}^{\Uc_1}=\left\{ Z_{\Vc,i}^{\Uc_1} : i\in [\Usf] \right\}$, where each coded key contains $\frac{\Lsf}{\Usf\left(\binom{\Ksf-1}{\Ssf-1}-\binom{\Ksf-1-\Usf}{\Ssf-1}\right)}$ symbols. We denote the sets in the collection $\left\{ \Vc\in \binom{[\Ksf]}{\Ssf} : \Vc\cap \Uc_1 \neq \emptyset \right\}$, by $\Vc_{\Uc_1,1}, \Vc_{\Uc_1,2},\ldots,\Vc_{\Uc_1,P}$, where with a slight abuse of notation $P$ represents the number of sets in the above collection. 
Thus the recovery task in~\eqref{eq:round 2 total task} could be expressed in the   matrix form   
\begin{align}
{\bf F}  \left[ \begin{array}{c}
Z_{\Vc_{\Uc_1,1}, 1}^{\Uc_1}\\
Z_{\Vc_{\Uc_1,2}, 1}^{\Uc_1}\\
\vdots \\
Z_{\Vc_{\Uc_1,P}, 1}^{\Uc_1}\\
Z_{\Vc_{\Uc_1,1}, 2}^{\Uc_1}\\
 \vdots\\
 Z_{\Vc_{\Uc_1,P}, \Usf}^{\Uc_1}
\end{array} \right]=
\left[ \begin{array}{c}
F_1\\
\vdots \\
 F_{\Usf \binom{\Ksf-1}{\Ssf-1} }
\end{array} \right]
\label{eq:matrix form second round task}
\end{align}
where (recall that  
$(\mathbf{M})_{m \times n}$ represents the dimension of matrix $\mathbf{M}$ is $m \times n$)
 \begin{align}
 {\bf F}   = \left[\begin{array}{c:c:c:c}
 ({\bf A})_{ \binom{\Ksf-1}{\Ssf-1} \times P}  & {\bf 0}_{\binom{\Ksf-1}{\Ssf-1} \times P}  & \cdots & {\bf 0}_{\binom{\Ksf-1}{\Ssf-1} \times P}   \\ \hdashline
{\bf 0}_{\binom{\Ksf-1}{\Ssf-1} \times P}  &  ({\bf A})_{ \binom{\Ksf-1}{\Ssf-1} \times P}   & \cdots & {\bf 0}_{\binom{\Ksf-1}{\Ssf-1} \times P}    \\ \hdashline 
 \vdots   & \vdots  &  \ddots& \vdots \\ \hdashline
{\bf 0}_{\binom{\Ksf-1}{\Ssf-1} \times P}  &   {\bf 0}_{\binom{\Ksf-1}{\Ssf-1} \times P}     & \cdots &  ({\bf A})_{ \binom{\Ksf-1}{\Ssf-1} \times P} 
 \end{array}
\right], \label{eq:constructed F}
 \end{align}
 and 
 \begin{align}
 ({\bf A})_{ \binom{\Ksf-1}{\Ssf-1} \times P} = \left[\begin{array}{c:c:c:c}
 a_{\Vc_{\Uc_1,1},1} &  a_{\Vc_{\Uc_1,2},1} & \cdots &  a_{\Vc_{\Uc_1,P},1} \\
  a_{\Vc_{\Uc_1,1},2} &  a_{\Vc_{\Uc_1,2},2} & \cdots &  a_{\Vc_{\Uc_1,P},2} \\
\vdots   & \vdots  &  \ddots& \vdots \\
a_{\Vc_{\Uc_1,1},\binom{\Ksf-1}{\Ssf-1}} &  a_{\Vc_{\Uc_1,2},\binom{\Ksf-1}{\Ssf-1}} & \cdots &  a_{\Vc_{\Uc_1,P},\binom{\Ksf-1}{\Ssf-1}}
  \end{array}
\right]. \label{eq:matrix A}
 \end{align}
Note that 
$$\left\{ F_{\binom{\Ksf-1}{\Ssf-1} (i-1) +j} : i\in [\Usf], j\in \left[\binom{\Ksf-1}{\Ssf-1}-\binom{\Ksf-1-\Usf}{\Ssf-1}+1 : \binom{\Ksf-1}{\Ssf-1} \right] \right\}
$$ is the set of linear combinations in~\eqref{eq:round 2 knows}, 
 which have already been recovered by the server in the first round.

Thus 
 we can let each user $k\in \Uc_1$ transmit 
\begin{align}
Y^{\Uc_1}_{k} =    {\bf S}_{k}    \begin{bmatrix}
F_1  \\
\vdots\\
F_{\Usf \binom{\Ksf-1}{\Ssf-1}}
\end{bmatrix}   ,
\label{eq:transmission of Y_k}
\end{align}
where ${\bf S}_{k}$  with $ \binom{\Ksf-1}{\Ssf-1}-\binom{\Ksf-1-\Usf}{\Ssf-1}$ rows  represents the matrix of second-round transmission by user $k$. The next step is to determine ${\bf S}_{k}$ satisfying that
\begin{itemize}
\item (c1) in the transmission by each user $k\in \Uc_1$, the coefficients of   $Z^{\Uc_1}_{\Vc,i}$ where $\Vc\in \binom{[\Ksf]\setminus\{k\}}{\Ssf}$ and $i\in [\Usf]$ are $0$, since user $k$ cannot compute such coded keys;
\item (c2) for any set $\Uc_2\subseteq \Uc_1$ where $|\Uc_2|=\Usf$, the matrix 
\begin{align}
\begin{bmatrix}
{\bf S}_{\Uc_{2,1}}  \\
\vdots\\
{\bf S}_{\Uc_{2,\Usf}}\\
\ev_{\Usf\binom{\Ksf-1}{\Ssf-1},\binom{\Ksf-1}{\Ssf-1}-\binom{\Ksf-1-\Usf}{\Ssf-1}+1}\\
\ev_{\Usf\binom{\Ksf-1}{\Ssf-1},\binom{\Ksf-1}{\Ssf-1}-\binom{\Ksf-1-\Usf}{\Ssf-1}+2}\\
\vdots\\
\ev_{\Usf\binom{\Ksf-1}{\Ssf-1},\binom{\Ksf-1}{\Ssf-1}}\\
\ev_{\Usf\binom{\Ksf-1}{\Ssf-1},\binom{\Ksf-1}{\Ssf-1}+\binom{\Ksf-1}{\Ssf-1}-\binom{\Ksf-1-\Usf}{\Ssf-1}+1}\\
\vdots\\
\ev_{\Usf\binom{\Ksf-1}{\Ssf-1},\binom{\Ksf-1}{\Ssf-1}+\binom{\Ksf-1}{\Ssf-1}}\\
\vdots\\
\ev_{\Usf\binom{\Ksf-1}{\Ssf-1},\Usf\binom{\Ksf-1}{\Ssf-1}}
\end{bmatrix}
\label{eq:decodability matrix}
\end{align}
 with dimension $\Usf\binom{\Ksf-1}{\Ssf-1} \times \Usf\binom{\Ksf-1}{\Ssf-1}$ is full rank. 
\end{itemize}
Note that (c1) guarantees in $Y^{\Uc_1}_{k}$, the transmitted linear combinations of the coded keys do not contain the coded keys which user $k$ cannot compute; (c2) guarantees that from  the second-round transmissions by any  $\Usf$ users in $\Uc_1$, the server can recover~\eqref{eq:matrix form second round task}  and then   the computation task $\sum_{k^{\prime}\in\Uc_1} W_{k^{\prime},j}$ for each $j\in \left[ \binom{\Ksf-1}{\Ssf-1}-\binom{\Ksf-1-\Usf}{\Ssf-1} \right]$.

Let us focus on one user $k\in [\Ksf]$ and   construct   ${\bf S}_{k}$; note that, our construction on ${\bf S}_{k}$ is independent of the value of $\Uc_1$. Denote   the sets  in $\binom{[\Ksf]\setminus\{k\}}{\Ssf}$   by  
$\overline{\Sc}_{k,1},\ldots,\overline{\Sc}_{k,\binom{\Ksf-1}{\Ssf}}$.
By the choice of the coefficient vectors, we have the following lemma, whose proof could be found in Appendix~\ref{sec:proof of Lemma IA}. 
\begin{lem}
\label{lem:IA}
The matrix
\begin{align}
\left[\av_{\overline{\Sc}_{k,1}}, \av_{\overline{\Sc}_{k,2}},\ldots, \av_{\overline{\Sc}_{k, \binom{\Ksf-1}{\Ssf}}}  \right] \label{eq:unknown matrix}
\end{align} 
with dimension $\binom{\Ksf-1}{\Ssf-1} \times \binom{\Ksf-1}{\Ssf}$, has rank equal to $\binom{\Ksf-2}{\Ssf-1}$ with high probability. 
\hfill $\square$  
\end{lem}
It can be seen from Lemma~\ref{lem:IA} that, by the choice of the coefficient vectors, the ``interferences'' to user $k$ (i.e., the coded keys which user $k$ cannot compute) are aligned. 
Thus with high probability the left null space of the matrix in~\eqref{eq:unknown matrix} contains $\binom{\Ksf-1}{\Ssf-1} - \binom{\Ksf-2}{\Ssf-1}= \binom{\Ksf-2}{\Ssf-2}$ linearly independent vectors, denoted by $\sv_{k,1},\ldots, \sv_{k, \binom{\Ksf-2}{\Ssf-2}}$, each with dimension   $1\times \binom{\Ksf-1}{\Ssf-1}$. 

Then considering the division of keys in the second-round transmission, we construct the following matrix with dimension $\Usf \binom{\Ksf-2}{\Ssf-2} \times \Usf \binom{\Ksf-1}{\Ssf-1}$, 
\begin{align}
{\bf S}^{\prime}_k = \left[\begin{array}{c:c:c:c}
 \begin{matrix}
\sv_{k,1} \\
\vdots\\
\sv_{k, \binom{\Ksf-2}{\Ssf-2}}
\end{matrix} &   \begin{matrix}
{\bf 0}_{1\times \binom{\Ksf-1}{\Ssf-1} } \\
\vdots\\
{\bf 0}_{1\times \binom{\Ksf-1}{\Ssf-1} }
\end{matrix}    & \begin{matrix}
\cdots \\
\ddots\\
\cdots
\end{matrix}    & \begin{matrix}
{\bf 0}_{1\times \binom{\Ksf-1}{\Ssf-1} } \\
\vdots\\
{\bf 0}_{1\times \binom{\Ksf-1}{\Ssf-1} }
\end{matrix}   \\ \hdashline
\begin{matrix}
{\bf 0}_{1\times \binom{\Ksf-1}{\Ssf-1} } \\
\vdots\\
{\bf 0}_{1\times \binom{\Ksf-1}{\Ssf-1} }
\end{matrix} &  \begin{matrix}
\sv_{k,1} \\
\vdots\\
\sv_{k, \binom{\Ksf-2}{\Ssf-2}}
\end{matrix}   & \begin{matrix}
\cdots \\
\ddots\\
\cdots
\end{matrix}  & \begin{matrix}
{\bf 0}_{1\times \binom{\Ksf-1}{\Ssf-1} } \\
\vdots\\
{\bf 0}_{1\times \binom{\Ksf-1}{\Ssf-1} }
\end{matrix}   \\ \hdashline 
 \vdots   & \vdots  &  \ddots& \vdots \\ \hdashline
\begin{matrix}
{\bf 0}_{1\times \binom{\Ksf-1}{\Ssf-1} } \\
\vdots\\
{\bf 0}_{1\times \binom{\Ksf-1}{\Ssf-1} }
\end{matrix}    &  \begin{matrix}
{\bf 0}_{1\times \binom{\Ksf-1}{\Ssf-1} } \\
\vdots\\
{\bf 0}_{1\times \binom{\Ksf-1}{\Ssf-1} }
\end{matrix}       & \begin{matrix}
\cdots \\
\ddots\\
\cdots
\end{matrix}  &  \begin{matrix}
\sv_{k,1} \\
\vdots\\
\sv_{k, \binom{\Ksf-2}{\Ssf-2}}
\end{matrix} 
 \end{array}
\right].
\label{eq:matrix k before RLC}
\end{align}
Finally, we let ${\bf S}_k $ be $\binom{\Ksf-1}{\Ssf-1}-\binom{\Ksf-1-\Usf}{\Ssf-1}$ random linear combinations of the rows in ${\bf S}^{\prime}_k$, where each coefficient in each linear combination is uniformly i.i.d. over $\mathbb{F}_{\qsf}$. This is possible since 
\begin{align}
\Usf \binom{\Ksf-2}{\Ssf-2} \geq \binom{\Ksf-1}{\Ssf-1} - \binom{\Ksf-1-\Usf}{\Ssf-1}, \label{eq:enough space}
\end{align}
whose proof could be found in Appendix~\ref{sec:proof of eq enough space}. Note that it will be also proved in Appendix~\ref{sec:proof of eq enough space} that the equality in~\eqref{eq:enough space} holds when $\Ksf=\Ssf$. 

By construction,   the columns in ${\bf S}^{\prime}_k {\bf A}$ corresponding to the coded keys $Z^{\Uc_1}_{\Vc,j}$ where $\Vc \in \binom{[\Ksf]\setminus \{k\}}{\Ssf}$,  $\Vc\cap \Uc_1 \neq \emptyset$, and $j\in [\Usf]$, are all ${\bf 0}_{\Usf \binom{\Ksf-2}{\Ssf-2} \times 1}$. Since the rows of ${\bf S}_k $ are linear combinations of the rows in ${\bf S}^{\prime}_k$,   the columns  in ${\bf S}_k {\bf A}$ corresponding to the coded keys $Z^{\Uc_1}_{\Vc,j}$ where $\Vc \in \binom{[\Ksf]\setminus \{k\}}{\Ssf}$,  $\Vc\cap \Uc_1 \neq \emptyset$, and $j\in [\Usf]$, are all ${\bf 0}_{\left(\binom{\Ksf-1}{\Ssf-1} - \binom{\Ksf-1-\Usf}{\Ssf-1}\right) \times 1}$. 
Thus (c1) is satisfied.

In Appendix~\ref{sec:proof of SZ lemma}, by using the Schwartz-Zippel lemma~\cite{Schwartz,Zippel,Demillo_Lipton}, we have the following lemma, which shows that (c2) is satisfied by the proposed second-round transmission.
\begin{lem}
\label{lem:SZ lemma}
For any set $\Uc_2\subseteq  [\Ksf]$ where $|\Uc_2|=\Usf$, the matrix in~\eqref{eq:decodability matrix} is full rank with high probability. 
\hfill $\square$  
\end{lem}

As a result, the decodability of the proposed scheme is proved.\footnote{\label{foot:whp}Lemma~\ref{lem:SZ lemma} shows that by randomly selecting some coefficients, the proposed secure aggregation is decodable with high probability; thus there must exist one choice of those coefficients such that the proposed scheme is decodable.}

For the security of the proposed scheme, we first provide an intuitive explanation. By~\eqref{eq:full rank constraint}, the server cannot get any information about $W_{1},\ldots,W_{\Ksf}$ from the first-round transmissions $X_{1},\ldots X_{\Ksf}$. In the second round, all the transmissions are in the linear space spanned by $F_1, \ldots
 F_{\Usf \binom{\Ksf-1}{\Ssf-1} }$, totally $\Usf \binom{\Ksf-1}{\Ssf-1} \frac{\Lsf}{\Usf \binom{\Ksf-1}{\Ssf-1}}= \Lsf$ symbols. Hence, 
by~\cite{shannonsecurity}, from the second round the server can get at most $\Lsf$ symbols information about $W_{1},\ldots,W_{\Ksf}$, which are exactly the $\Lsf$ symbols in the   computation task $\sum_{k\in\Uc_1} W_{k,j}$
by the decodability proof. Hence, 
 the server cannot get any information about $W_{1},\ldots,W_{\Ksf}$  except the computation task. Following the above intuitive explanation, the information theoretic security proof 
 could be directly obtained as in~\cite[Appendix C]{groupwisekey2022wan}. 
 

\begin{rem}
\label{rem:also work for}
The proposed secure aggregation scheme in this section can also work for the case $\Ssf> \Ksf-\Usf$. In this case, $\Rsf_1=\frac{\binom{\Ksf-1}{\Ssf-1}}{\binom{\Ksf-1}{\Ssf-1}-\binom{\Ksf-1-\Usf}{\Ssf-1}}=1$ and each input vector is divided into $ \binom{\Ksf-1}{\Ssf-1}-\binom{\Ksf-1-\Usf}{\Ssf-1} =  \binom{\Ksf-1}{\Ssf-1}$ non-overlapping and equal-length pieces.
In the first round  transmission, we only transmit  the coded messages in~\eqref{eq:first part first round}, while the coded messages in~\eqref{eq:second part first round} does not exist since $\binom{\Ksf-1-\Usf}{\Ssf-1}=0$. In other words, since the number of users knowing each key is larger than the maximal number of dropped users, in the first round we do not need to transmit coded messages in~\eqref{eq:second part first round} which are only composed of keys. In addition, in the second round transmission, the proposed scheme also works with the optimal communication rate $\Rsf_2=1/\Usf$, while the decodability and security constraints are both satisfied. 

For   the case $\Ssf> \Ksf-\Usf$,  compared to the  secure aggregation scheme in~\cite{groupwisekey2022wan} with the optimal communication rates,  the proposed secure aggregation scheme in this paper achieves the same optimal communication rates. However, the proposed scheme 
requires all the $\binom{\Ksf}{\Ssf}$ keys each of which is shared by a different set of $\Ssf$ users and has $\frac{\Ssf}{\binom{\Ksf-1}{\Ssf-1}}$ symbols; the number of keys required by the scheme in~\cite{groupwisekey2022wan} is at most $\Oc(\Ksf^2)$, where each key is shared by $\Ssf$ users and has   
$(\Ksf-\Usf+1)\Lsf/\Usf$ symbols.
\hfill $\square$ 
\end{rem}

\section{Experimental Results}
\label{sec:experiment}
We implement our proposed secure aggregation scheme  in Python3.11 by using the MPI4py library over the Tencent Cloud, which is then compared to the original secure aggregation scheme in \cite{bonawitz2017practical} (referred to as SecAgg).
Our comparison focuses   on the model aggregation times of the proposed protocol  and SecAgg, by assuming the offline keys have been already shared. 

\begin{figure}
    \centering
        \includegraphics[scale=0.7]{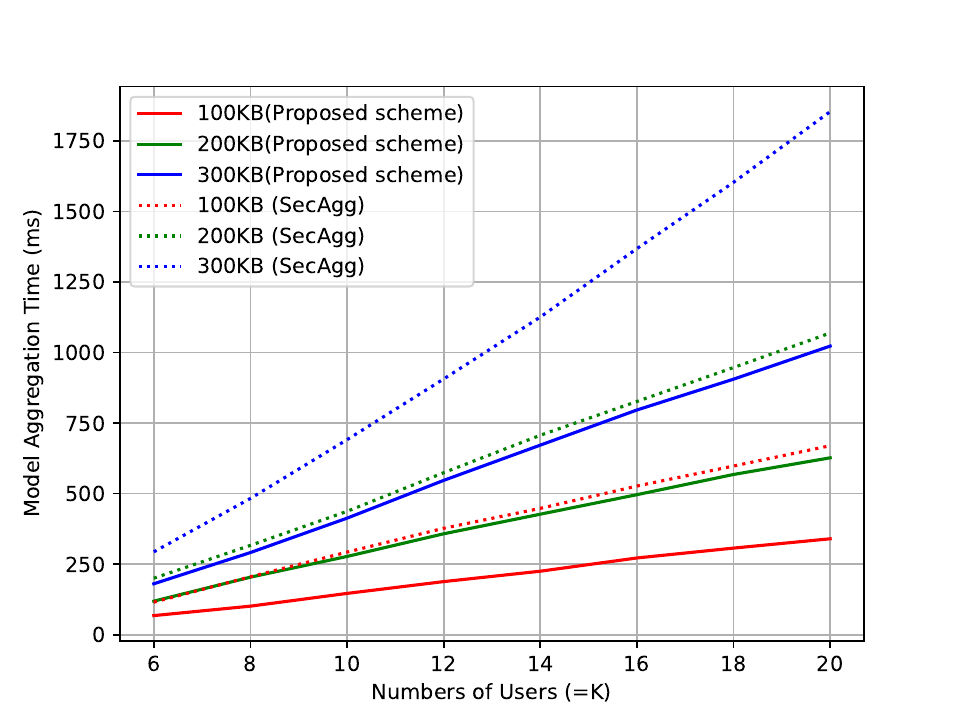}
    \caption{\small Model aggregation time of the proposed scheme and \texttt{SecAgg} for $\Usf =\left\lfloor  (\Ksf + 1)/2 \right\rfloor  $ and $\Ssf=\Ksf-\Usf$.}
    \label{fig:the comparison of model aggregation time}
\end{figure}

{\bf Tencent Cloud Setup}.
We choose Tencent Cloud instances, specifically \verb"S6.LARGE16" and \verb"S6.MEDIUM2". Of these, one \verb"S6.LARGE16" instance plays the role of the server and all the users. These Tencent Cloud instances are equipped with Intel Xeon Ice Lake processors running at a base clock speed of 2.7 GHz with a turbo frequency of 3.3 GHz. All instances used in our experiment are identical in terms of computing power, memory and network resources. The communication speed between the server and the users is a fast 100MB/s.
To generate our input vectors, we set the field size $\qsf$ to $7$ and generate vectors that are uniformly and independently distributed over $\mathbb{F}_7$. We also consider three different sizes for each input vector: 100KB, 200KB, and 300KB, following the suggestions in \cite{bonawitz2017practical}. In the system configuration represented by $(\Ksf,\Usf,\Ssf)$, we use Monte Carlo methods with $100$ samples and then average the resulting times over these $100$ samples.

{\bf Proposed scheme v.s. \texttt{SecAgg}.} 
We compare the model aggregation times (including encryption, transmission, decryption times) of  our proposed scheme and SecAgg, for  $\Usf =\left\lfloor  (\Ksf + 1)/2 \right\rfloor  $ and $\Ssf=\Ksf-\Usf$, as shown in Fig.~\ref{fig:the comparison of model aggregation time}. We can see that the proposed scheme outperforms SecAgg by  reducing model aggregation time, where the reduction percent  ranges from   $29.7\%$ to   $67.2\%$. This improvement coincides with the theoretical perspective that the proposed scheme  achieves the optimal communication cost during the model aggregation phase, while SecAgg does not.

\section{Conclusions}
\label{sec:conclusion}
In this paper, we aimed to minimize  the communication rates in the two-round transmission of the model aggregation phase, for the information theoretic secure aggregation problem with uncoded groupwise keys. While preserving the security on the users' local data, the secure aggregation scheme should also be able to tolerate user dropouts.  By 
 proposing a new and tight converse bound    coinciding with the achievable rates by the new  secure aggregation scheme based on interference alignment, we fully characterized the region of all possible rates tuples for the considered problem.  Ongoing work includes further reducing the requirement on the size of keys  and extending the secure aggregation scheme with uncoded groupwise keys to the scenario of clustered federated learning.

\appendices
\section{Proof of Lemma~\ref{lem:crucial lemma on transform} }
\label{sec:proof of crucial lemma on transform}
Consider one set $\Vc \in \binom{[\Ksf]}{\Ssf}$ and one $k\in [\Ksf]\setminus \Vc$. Recall that $n_{\Vc,k}$ represents the number of elements in $\Vc$ which are smaller than $k$.
 
When $k=1$, we have $n_{\Vc,1}=0$ and thus the equation to be proved, ~\eqref{eq:general tranform eq}, becomes 
\begin{align}
\av_{\Vc}= \sum_{i_1\in [\Ssf]}(-1)^{i_1-1}  \av_{\Vc\setminus \{\Vc(i_1)\} \cup \{1\}},
\end{align} 
which is exactly~\eqref{eq:construction of other av} and thus is proved.  

When $1\in \Vc$ and $k\neq 1$, assume that $\Vc \setminus \{1\} = \Vc^{\prime} $ where $|\Vc^{\prime}|=\Ssf-1$. Hence,~\eqref{eq:general tranform eq} becomes
\begin{subequations}
\begin{align}
& \av_{\Vc^{\prime}  \cup \{1\}}=   \sum_{i_1\in [n_{\Vc,k}+1:\Ssf]}(-1)^{i_1-n_{\Vc,k}-1}  \av_{\Vc\setminus \{\Vc(i_1)\} \cup \{k\}}   + 
 (-1)^{n_{\Vc,k}+1} \av_{\Vc^{\prime} \cup \{k\}} + 
  \sum_{i_2\in [2:n_{\Vc,k}] } (-1)^{n_{\Vc,k}+i_2} \av_{\Vc\setminus \{\Vc(i_2)\} \cup \{k\}},   \label{eq:general transform eq becomes} \\
  & \Longleftrightarrow 
  (-1)^{n_{\Vc,k}} \av_{\Vc^{\prime} \cup \{k\}} =  \sum_{i_1\in [n_{\Vc,k}+1:\Ssf]}(-1)^{i_1-n_{\Vc,k}-1}  \av_{\Vc\setminus \{\Vc(i_1)\} \cup \{k\}}  + 
  \sum_{i_2\in [2:n_{\Vc,k}] } (-1)^{n_{\Vc,k}+i_2} \av_{\Vc\setminus \{\Vc(i_2)\} \cup \{k\}} -\av_{\Vc^{\prime}  \cup \{1\}}, \\
&  \Longleftrightarrow  
\av_{\Vc^{\prime} \cup \{k\}}=  \sum_{i_2\in [2:n_{\Vc,k}] } (-1)^{i_2} \av_{\Vc\setminus \{\Vc(i_2)\} \cup \{k\}} 
+ (-1)^{n_{\Vc,k}+1} \av_{\Vc^{\prime}  \cup \{1\}} + \sum_{i_1\in [n_{\Vc,k}+1:\Ssf]}(-1)^{i_1-1}  \av_{\Vc\setminus \{\Vc(i_1)\} \cup \{k\}}, \\
& \Longleftrightarrow  
\av_{\Vc^{\prime} \cup \{k\}}=  \sum_{i_4\in [1:n_{\Vc,k}-1] } (-1)^{i_4-1} \av_{\Vc \cup \{k\} \setminus \{\Vc^{\prime}(i_4)\} } 
+ (-1)^{n_{\Vc,k}-1} \av_{\Vc^{\prime}  \cup \{1\}} + \sum_{i_3\in [n_{\Vc,k}:\Ssf-1]}(-1)^{i_3}  \av_{\Vc \cup \{k\} \setminus \{\Vc^{\prime}(i_3)\} } ,
  \label{eq:general transform eq to be proved}
\end{align} 
\end{subequations}
where~\eqref{eq:general transform eq becomes} follows since $\Vc(1)=1$ and~\eqref{eq:general transform eq to be proved} follows since we let $i_3=i_1-1$ and $i_4=i_2-1$. 
It can be seen that~\eqref{eq:general transform eq to be proved} can be  derived  by directly   expanding  $\av_{\Vc^{\prime} \cup \{k\}}$ according to~\eqref{eq:construction of other av}.

Finally, we consider the most involved case where $k\neq 1$ and $1\notin \Vc$.  We first expand the LHS of~\eqref{eq:general tranform eq} as in~\eqref{eq:construction of other av}; that is, 
\begin{align}
 \av_{\Vc}= \sum_{i\in [\Ssf]} (-1)^{i-1}  \av_{\Vc\setminus \{\Vc(i)\} \cup \{1\}}. \label{eq:write again av}
\end{align}
Then for each $i\in [\Ssf]$, we will show in the following that   the RHS of~\eqref{eq:general tranform eq} also contains the term $(-1)^{i-1} \av_{\Vc\setminus \{\Vc(i)\} \cup \{1\}}$. 
We also expand each term on the RHS of~\eqref{eq:general tranform eq} by using~\eqref{eq:construction of other av}.   Note that $\av_{\Vc\setminus \{\Vc(i)\} \cup \{1\}}$ can only appear in 
$\av_{\Vc\setminus \{\Vc(i)\} \cup \{k\}}$. 
We consider two cases:
\begin{itemize}
\item $\Vc(i)>k$. 
The coefficient of $\av_{\Vc\setminus \{\Vc(i)\} \cup \{k\}}$ on the RHS of~\eqref{eq:general tranform eq} is $(-1)^{i-n_{\Vc,k}-1}$. We expand  $\av_{\Vc\setminus \{\Vc(i)\} \cup \{k\}}$ by using~\eqref{eq:construction of other av}. Since the number of  elements in $\Vc$ smaller than $k$ is $n_{\Vc,k}$ and $\Vc(i)>k$, the number of elements in $\Vc\setminus \{\Vc(i)\} \cup \{k\}$ smaller than $k$ is also  $n_{\Vc,k}$; thus  by using~\eqref{eq:construction of other av}, 
the coefficient of  $\av_{\Vc\setminus \{\Vc(i)\} \cup \{1\}}$ in $\av_{\Vc\setminus \{\Vc(i)\} \cup \{k\}}$ is $(-1)^{n_{\Vc,k}}.$ Hence, the   coefficient of  $\av_{\Vc\setminus \{\Vc(i)\} \cup \{1\}}$ on the RHS of~\eqref{eq:general tranform eq} is $(-1)^{i-n_{\Vc,k}-1} (-1)^{n_{\Vc,k}}=(-1)^{i-1}$.
\item $\Vc(i)<k$. The coefficient of $\av_{\Vc\setminus \{\Vc(i)\} \cup \{k\}}$ on the RHS of~\eqref{eq:general tranform eq} is $(-1)^{n_{\Vc,k}+i}$. We expand  $\av_{\Vc\setminus \{\Vc(i)\} \cup \{k\}}$ by using~\eqref{eq:construction of other av}. Since the number of  elements in $\Vc$ smaller than $k$ is $n_{\Vc,k}$ and $\Vc(i)<k$, the number of elements in $\Vc\setminus \{\Vc(i)\} \cup \{k\}$ smaller than $k$ is    $n_{\Vc,k}-1$; thus  by using~\eqref{eq:construction of other av}, 
the coefficient of  $\av_{\Vc\setminus \{\Vc(i)\} \cup \{1\}}$ in $\av_{\Vc\setminus \{\Vc(i)\} \cup \{k\}}$ is $(-1)^{n_{\Vc,k}-1}.$ Hence, the   coefficient of  $\av_{\Vc\setminus \{\Vc(i)\} \cup \{1\}}$ on the RHS of~\eqref{eq:general tranform eq} is $(-1)^{n_{\Vc,k}+i} (-1)^{n_{\Vc,k}-1}=(-1)^{i-1}$.
\end{itemize}

After expanding each  $\av_{\Vc\setminus \{\Vc(i)\} \cup \{k\}} $  where $i \in [\Ssf]$
on the RHS of~\eqref{eq:general tranform eq} by using~\eqref{eq:construction of other av},  the RHS of~\eqref{eq:general tranform eq} may only contain $\av_{\Vc\setminus \{\Vc(i)\} \cup \{1\}}$ where $i\in [\Ssf]$ and $\av_{\Vc\setminus \{\Vc(i^{\prime}),\Vc(i^{\prime\prime})\} \cup \{1,k\}}$ where $1\leq i^{\prime}<i^{\prime\prime} \leq \Ssf$. 
Next we will prove that the coefficient  of each  term  $\av_{\Vc\setminus \{\Vc(i^{\prime}),\Vc(i^{\prime\prime})\} \cup \{1,k\}}$  where $1\leq i^{\prime}<i^{\prime\prime} \leq \Ssf$, is $0$.

$\av_{\Vc\setminus \{\Vc(i^{\prime}),\Vc(i^{\prime\prime})\} \cup \{1,k\}}$ appears in the expansions of $\av_{\Vc\setminus \{\Vc(i^{\prime})\} \cup \{k\}}$ and $\av_{\Vc\setminus \{\Vc(i^{\prime\prime})\} \cup \{k\}}$. 
We consider three cases:
\begin{itemize}
\item $\Vc(i^{\prime})<\Vc(i^{\prime\prime})<k$. The coefficient of $\av_{\Vc\setminus \{\Vc(i^{\prime})\} \cup \{k\}}$ on the RHS of~\eqref{eq:general tranform eq} is $(-1)^{n_{\Vc,k}+i^{\prime}}$; the  coefficient of
$\av_{\Vc\setminus \{\Vc(i^{\prime}),\Vc(i^{\prime\prime})\} \cup \{1,k\}}$ in the expansion of $\av_{\Vc\setminus \{\Vc(i^{\prime})\} \cup \{k\}}$ is $(-1)^{i^{\prime\prime}-1-1}=(-1)^{i^{\prime\prime}}$, since the number of elements in $\Vc\setminus \{\Vc(i^{\prime})\} \cup \{k\}$ smaller than $\Vc(i^{\prime\prime})$ is $i^{\prime\prime}-1-1$. 
Similarly, the coefficient of $\av_{\Vc\setminus \{\Vc(i^{\prime\prime})\} \cup \{k\}}$ on the RHS of~\eqref{eq:general tranform eq} is $(-1)^{n_{\Vc,k}+i^{\prime\prime}}$; the  coefficient of
$\av_{\Vc\setminus \{\Vc(i^{\prime}),\Vc(i^{\prime\prime})\} \cup \{1,k\}}$ in the expansion of $\av_{\Vc\setminus \{\Vc(i^{\prime\prime})\} \cup \{k\}}$ is $(-1)^{i^{\prime}-1}$, since the number of elements in $\Vc\setminus \{\Vc(i^{\prime\prime})\} \cup \{k\}$ smaller than $\Vc(i^{\prime})$ is $i^{\prime}-1$.  
Hence, the coefficient of $\av_{\Vc\setminus \{\Vc(i^{\prime}),\Vc(i^{\prime\prime})\} \cup \{1,k\}}$ on the RHS of~\eqref{eq:general tranform eq}  is 
$$
(-1)^{n_{\Vc,k}+i^{\prime}} (-1)^{i^{\prime\prime}} + (-1)^{n_{\Vc,k}+i^{\prime\prime}}(-1)^{i^{\prime}-1}=0. 
$$
\item $k<\Vc(i^{\prime})<\Vc(i^{\prime\prime})$.  
The coefficient of $\av_{\Vc\setminus \{\Vc(i^{\prime})\} \cup \{k\}}$ on the RHS of~\eqref{eq:general tranform eq} is $(-1)^{i^{\prime}-n_{\Vc,k}-1}$; the  coefficient of
$\av_{\Vc\setminus \{\Vc(i^{\prime}),\Vc(i^{\prime\prime})\} \cup \{1,k\}}$ in the expansion of $\av_{\Vc\setminus \{\Vc(i^{\prime})\} \cup \{k\}}$ is $(-1)^{i^{\prime\prime}-1}$, since the number of elements in $\Vc\setminus \{\Vc(i^{\prime})\} \cup \{k\}$ smaller than $\Vc(i^{\prime\prime})$ is $i^{\prime\prime}-1$.  
Similarly, the coefficient of $\av_{\Vc\setminus \{\Vc(i^{\prime\prime})\} \cup \{k\}}$ on the RHS of~\eqref{eq:general tranform eq} is $(-1)^{i^{\prime\prime}-n_{\Vc,k}-1}$; the  coefficient of
$\av_{\Vc\setminus \{\Vc(i^{\prime}),\Vc(i^{\prime\prime})\} \cup \{1,k\}}$ in the expansion of $\av_{\Vc\setminus \{\Vc(i^{\prime\prime})\} \cup \{k\}}$ is $(-1)^{i^{\prime}}$, since the number of elements in $\Vc\setminus \{\Vc(i^{\prime\prime})\} \cup \{k\}$ smaller than $\Vc(i^{\prime})$ is $i^{\prime}$.  
Hence, the coefficient of $\av_{\Vc\setminus \{\Vc(i^{\prime}),\Vc(i^{\prime\prime})\} \cup \{1,k\}}$ on the RHS of~\eqref{eq:general tranform eq}  is 
$$
(-1)^{i^{\prime}-n_{\Vc,k}-1} (-1)^{i^{\prime\prime}-1}   + (-1)^{i^{\prime\prime}-n_{\Vc,k}-1} (-1)^{i^{\prime}} =0.
$$
\item $\Vc(i^{\prime})<k<\Vc(i^{\prime\prime})$. 
 The coefficient of $\av_{\Vc\setminus \{\Vc(i^{\prime})\} \cup \{k\}}$ on the RHS of~\eqref{eq:general tranform eq} is $(-1)^{n_{\Vc,k}+i^{\prime}}$; the  coefficient of
$\av_{\Vc\setminus \{\Vc(i^{\prime}),\Vc(i^{\prime\prime})\} \cup \{1,k\}}$ in the expansion of $\av_{\Vc\setminus \{\Vc(i^{\prime})\} \cup \{k\}}$ is $(-1)^{i^{\prime\prime}-1}$, since the number of elements in $\Vc\setminus \{\Vc(i^{\prime})\} \cup \{k\}$ smaller than $\Vc(i^{\prime\prime})$ is $i^{\prime\prime}-1$. 
In addition, the coefficient of $\av_{\Vc\setminus \{\Vc(i^{\prime\prime})\} \cup \{k\}}$ on the RHS of~\eqref{eq:general tranform eq} is $(-1)^{i^{\prime\prime}-n_{\Vc,k}-1}$; the  coefficient of
$\av_{\Vc\setminus \{\Vc(i^{\prime}),\Vc(i^{\prime\prime})\} \cup \{1,k\}}$ in the expansion of $\av_{\Vc\setminus \{\Vc(i^{\prime\prime})\} \cup \{k\}}$ is $(-1)^{i^{\prime}-1}$, since the number of elements in $\Vc\setminus \{\Vc(i^{\prime\prime})\} \cup \{k\}$ smaller than $\Vc(i^{\prime})$ is $i^{\prime}-1$.  
Hence, the coefficient of $\av_{\Vc\setminus \{\Vc(i^{\prime}),\Vc(i^{\prime\prime})\} \cup \{1,k\}}$ on the RHS of~\eqref{eq:general tranform eq}  is 
$$
(-1)^{n_{\Vc,k}+i^{\prime}} (-1)^{i^{\prime\prime}-1} + (-1)^{i^{\prime\prime}-n_{\Vc,k}-1} (-1)^{i^{\prime}-1}=0. 
$$
\end{itemize}
As a result, we proved~\eqref{eq:general tranform eq}.  

\section{Proof of Lemma~\ref{lem:IA}}
\label{sec:proof of Lemma IA}
For each $k\in [\Ksf]$, we want to  prove that the matrix in~\eqref{eq:unknown matrix}, which is 
\begin{align}
\left[\av_{\overline{\Sc}_{k,1}}, \av_{\overline{\Sc}_{k,2}},\ldots, \av_{\overline{\Sc}_{k,\binom{\Ksf-1}{\Ssf}}}  \right], \label{eq:rank of unknown matrix}
\end{align}
  has rank   $\binom{\Ksf-2}{\Ssf-1}$ with high probability, where  $\av_{\overline{\Sc}_{k,1}}, \av_{\overline{\Sc}_{k,2}},\ldots, \av_{\overline{\Sc}_{k,\binom{\Ksf-1}{\Ssf}}}  $ denote the vectors $\Vc\in \binom{[\Ksf]\setminus \{k\} }{\Ssf}$. 

We select one user $k^{\prime} \in [\Ksf]\setminus \{k\}$. Denote the sets $\Vc\in \binom{[\Ksf]\setminus \{k\} }{\Ssf}$ where $k^{\prime}\in  \Vc$ by  $ \overline{\Sc}_{k,k^{\prime},1},\ldots, \overline{\Sc}_{k,k^{\prime},\binom{\Ksf-2}{\Ssf-1}}$. 
 It can be seen from~\eqref{eq:construction of other av} that, for each $\Vc\in \binom{[\Ksf]\setminus\{k,k^{\prime}\}}{\Ssf}$ we can re-construct  the vector $\av_{\Vc}$     by a linear combination of the  vectors 
 $\av_{\Vc\setminus \{k^{\prime\prime}\} \cup \{k^{\prime}\}}$ where $k^{\prime\prime}\in \Vc$. Hence, all the vectors $\av_{\overline{\Sc}_{k,1}}, \av_{\overline{\Sc}_{k,2}},\ldots, \av_{\overline{\Sc}_{k,\binom{\Ksf-1}{\Ssf}}}  $ are linear combinations of $ \overline{\Sc}_{k,k^{\prime},1},\ldots, \overline{\Sc}_{k,k^{\prime},\binom{\Ksf-2}{\Ssf-1}}$. Hence, the rank of the matrix in~\eqref{eq:rank of unknown matrix} is equal to the rank of 
 \begin{align}
 \left[  \overline{\Sc}_{k,k^{\prime},1},\ldots, \overline{\Sc}_{k,k^{\prime},\binom{\Ksf-2}{\Ssf-1}} \right]. \label{eq:basic rank unknown}
 \end{align}

By construction, all the vectors $\av_{\Vc}$ where $\Vc\in \binom{[\Ksf]}{\Ssf}$ are located in the linear space spanned by $\av_{\Vc_1}$ where $\Vc_1 \in\binom{[\Ksf]}{\Ssf}$ and $1\in \Vc_1$. Since each vector $\av_{\Vc_1}$ where $\Vc_1 \in\binom{[\Ksf]}{\Ssf}$ and $1\in \Vc_1$, is uniform and i.i.d. over $\mathbb{F}^{\binom{\Ksf-1}{\Ssf-1}}_{\qsf}$ with large enough $\qsf$, the above linear space has dimension $\binom{\Ksf-1}{\Ssf-1}$ with high probability. In addition by~\eqref{eq:construction of other av}, the vectors $\av_{\Vc_2}$ where $\Vc_2 \in\binom{[\Ksf]}{\Ssf}$ and $k^{\prime}\in \Vc_2$ can re-construct each vector in this linear space. Hence, the  $\binom{\Ksf-1}{\Ssf-1}$ vectors $\Vc_2 \in\binom{[\Ksf]}{\Ssf}$ and $k^{\prime}\in \Vc_2$ are linearly independent with high probability. Hence, we proved that the matrix in~\eqref{eq:basic rank unknown} is 
full rank with high probability, with rank $\binom{\Ksf-2}{\Ssf-1}$. As a result, we proved that the rank of the matrix in~\eqref{eq:rank of unknown matrix} is $\binom{\Ksf-2}{\Ssf-1}$ with high probability.

\section{Proof of~\eqref{eq:enough space}}
\label{sec:proof of eq enough space}
We prove~\eqref{eq:enough space} by induction. 

First consider the case $\Usf=1$. We need to prove 
\begin{align}
  \binom{\Ksf-2}{\Ssf-2} \geq \binom{\Ksf-1}{\Ssf-1} - \binom{\Ksf-2}{\Ssf-1}, \label{eq:U=2}
\end{align}
which directly holds from Pascal's triangle $ \binom{\Ksf-2}{\Ssf-2} = \binom{\Ksf-1}{\Ssf-1} - \binom{\Ksf-2}{\Ssf-1}$.

Then for any $\Usf \in [i]$, we assume that 
\begin{align}
i \binom{\Ksf-2}{\Ssf-2} \geq \binom{\Ksf-1}{\Ssf-1} - \binom{\Ksf-1-i}{\Ssf-1}  \label{eq:assume U holds}
\end{align} 
holds, and will prove 
\begin{align}
(i+1) \binom{\Ksf-2}{\Ssf-2} \geq \binom{\Ksf-1}{\Ssf-1} - \binom{\Ksf-i-2}{\Ssf-1}.  \label{eq:induction to prove}
\end{align}

Note that 
\begin{align}
\binom{\Ksf-2}{\Ssf-2}>\binom{\Ksf-i-2}{\Ssf-2}= \binom{\Ksf-i-1}{\Ssf-1}  -\binom{\Ksf-i-2}{\Ssf-1} . \label{eq:i+1 step}
\end{align}
By summing~\eqref{eq:assume U holds} and~\eqref{eq:i+1 step}, we can prove~\eqref{eq:induction to prove}. 

\section{Proof of Lemma~\ref{lem:SZ lemma}}
\label{sec:proof of SZ lemma}
Consider one set $\Uc_2 \subseteq [\Ksf]$ where $|\Uc_2|=\Usf$.  We want to prove that the matrix in~\eqref{eq:decodability matrix}
  with dimension $\Usf\binom{\Ksf-1}{\Ssf-1} \times \Usf\binom{\Ksf-1}{\Ssf-1}$ is full rank with high probability; i.e., the determinant of the matrix in~\eqref{eq:decodability matrix} is not zero with high probability. 
Note that   the determinant  could be written as $D_{\Ac}= \frac{P_{\Ac}}{Q_{\Ac}}$, where $P_{\Ac}$ and $Q_{\Ac}$ are multivariate polynomials whose variables are the elements in  $\av_{\Vc}$ where $\Vc \in \binom{[\Ksf]}{\Ssf}, 1\in \Vc$ and  the coefficients in the   $\binom{\Ksf-1}{\Ssf-1}-\binom{\Ksf-1-\Usf}{\Ssf-1}$ random linear combinations of the rows in ${\bf S}^{\prime}_k$ for each $k\in \Uc_2$.
Since the matrix in~\eqref{eq:decodability matrix} exists with high probability by Lemma~\ref{lem:IA} and~\eqref{eq:enough space}, $Q_{\Ac}$ is not zero with high probability. Hence, it remains to prove that $P_{\Ac}$ is not 
zero with high probability neither. 
Since  the variables in $P_{\Ac}$ are 
   uniform and i.i.d. over $\mathbb{F}_{\qsf}$ where   $\qsf$ is large enough,
by the  Schwartz-Zippel Lemma~\cite{Schwartz,Zippel,Demillo_Lipton}, if   the multivariate polynomial  $P_{\Ac}$ is   non-zero      (i.e., a multivariate polynomial whose coefficients are not all $0$), the probability that $P_{\Ac}$ is equal to $0$ over all possible realizations of variables   goes to $0$ when $\qsf$ goes to infinity, and thus 
  the matrix in~\eqref{eq:decodability matrix} is full rank with high probability. 
 So in the following, we need to show that $P_{\Ac}$ is a   non-zero polynomial; i.e., we want to find out one realization of the variables in $P_{\Ac}$, such that the matrix in~\eqref{eq:decodability matrix} exists and is full rank (in this way, $P_{\Ac}=D_{\Ac} Q_{\Ac}$ is not zero). 
 
 We pick one integer $u \in [\Ksf]\setminus \Uc_2$.  We first select the elements in  $\av_{\Vc}$ where $\Vc \in \binom{[\Ksf]}{\Ssf}, 1\in \Vc$. Note that the dimension of $\av_{\Vc}$ is $\binom{\Ksf-1}{\Ssf-1}$. 
  We want to let $[\av_{\Vc}:\Vc \in \binom{[\Ksf]}{\Ssf}, u\in \Vc]$ be an identity matrix with dimension $\binom{\Ksf-1}{\Ssf-1} \times \binom{\Ksf-1}{\Ssf-1}$. More precisely,  
  we define a collection of sets $\Cc_1= \left\{\Vc\in \binom{[\Ksf]}{\Ssf}: u\in \Vc, \Vc\cap \Uc_2\neq \emptyset \right\} $ and sort its sets as $\Cc_{1,1},\ldots, \Cc_{1,\binom{\Ksf-1}{\Ssf-1}-\binom{\Ksf-1-\Usf}{\Ssf-1} }$.  Let 
  \begin{align}
  \av_{\Cc_{1,i}}= \ev_{\binom{\Ksf-1}{\Ssf-1},i}, \ \forall i\in \left[ \binom{\Ksf-1}{\Ssf-1}-\binom{\Ksf-1-\Usf}{\Ssf-1} \right]. \label{eq:unit vector C1i}
  \end{align}
  In addition, we also  
define a collection of sets $\Cc_2= \left\{\Vc\in \binom{[\Ksf]}{\Ssf}: u\in \Vc, \Vc\cap \Uc_2=\emptyset \right\} $ and sort its sets as $\Cc_{2,1},\ldots, \Cc_{2,\binom{\Ksf-1-\Usf}{\Ssf-1} }$. Let 
\begin{align}
\av_{\Cc_{2,i}}= \ev_{\binom{\Ksf-1}{\Ssf-1},\binom{\Ksf-1}{\Ssf-1}-\binom{\Ksf-1-\Usf}{\Ssf-1}+i}, \ \forall i\in \left[ \binom{\Ksf-1-\Usf}{\Ssf-1}\right]. \label{eq:unit vector C2i}
\end{align}
There must exist a choice of $\av_{\Vc}$ where $\Vc \in \binom{[\Ksf]}{\Ssf}, 1\in \Vc$ which leads~\eqref{eq:unit vector C1i} and~\eqref{eq:unit vector C2i}. This is because by Lemma~\ref{lem:crucial lemma on transform}, we can obtain $\av_{\Vc}$ where $\Vc \in \binom{[\Ksf]}{\Ssf}, 1\in \Vc$ from the vectors $\av_{\Vc^{\prime}}$ where $\Vc^{\prime} \in \binom{[\Ksf]}{\Ssf}, u\in \Vc$. Hence, those resulting vectors $\av_{\Vc}$ where $\Vc \in \binom{[\Ksf]}{\Ssf}, 1\in \Vc$ can in turn lead~\eqref{eq:unit vector C1i} and~\eqref{eq:unit vector C2i}.

Focus on each user $k\in \Uc_2$.
Recall that the sets  in $\binom{[\Ksf]\setminus\{k\}}{\Ssf}$  are denoted by  $\overline{\Sc}_{k,1},\ldots,\overline{\Sc}_{k,\binom{\Ksf-1}{\Ssf}}$.
By Lemma~\ref{lem:crucial lemma on transform}, $\av_{\overline{\Sc}_{k,1}}, \av_{\overline{\Sc}_{k,2}},\ldots, \av_{\overline{\Sc}_{k,\binom{\Ksf-1}{\Ssf}}} $ are in the linear space spanned by 
$\av_{\Vc}$ where $\Vc \in \binom{[\Ksf]\setminus \{k\}}{\Ssf}$ and $u\in \Vc$. 
 Hence, the rank of the matrix $[\av_{\overline{\Sc}_{k,1}}, \av_{\overline{\Sc}_{k,2}},\ldots, \av_{\overline{\Sc}_{k,\binom{\Ksf-1}{\Ssf}}} ]$ is $\binom{\Ksf-2}{\Ssf-1}$; thus this matrix with dimension $\binom{\Ksf-1}{\Ssf-1} \times \binom{\Ksf-1}{\Ssf}$  contains $\binom{\Ksf-2}{\Ssf-2}$ linearly independent left null vectors, which are $\av_{\Vc}$ where $\Vc\in \binom{[\Ksf]}{\Ssf}$ and $\{u,k\} \subseteq \Vc$. 
 Let  ${\bf s}_{k,1}, \ldots, {\bf s}_{k,\binom{\Ksf-2}{\Ssf-2}}$ be these $\binom{\Ksf-2}{\Ssf-2}$ vectors, we can obtain ${\bf S}^{\prime}_k$ in~\eqref{eq:matrix k before RLC}.
 Since ${\bf S}^{\prime}_k$ exists, ${\bf S}_k$, which are $\binom{\Ksf-1}{\Ssf-1}-\binom{\Ksf-1-\Usf}{\Ssf-1}$ random linear combinations of the rows in ${\bf S}^{\prime}_k$,  also exists and thus the matrix in~\eqref{eq:decodability matrix} exists. 

Let us then select the coefficients in the   $\binom{\Ksf-1}{\Ssf-1}-\binom{\Ksf-1-\Usf}{\Ssf-1}$ random linear combinations of the $\Usf \binom{\Ksf-2}{\Ssf-2}$ rows in ${\bf S}^{\prime}_k$ for each $k\in \Uc_2$. 
More precisely, we directly select $\binom{\Ksf-1}{\Ssf-1}-\binom{\Ksf-1-\Usf}{\Ssf-1}$ rows from ${\bf S}^{\prime}_k$ to compose ${\bf S}_k$; i.e., in each linear combination, the coefficient vector contains   $\Usf \binom{\Ksf-2}{\Ssf-2}-1$ zeros and $1$ one. 
Hence, ${\bf S}_k$ contains $\binom{\Ksf-1}{\Ssf-1}-\binom{\Ksf-1-\Usf}{\Ssf-1}$ unit vectors, where these vectors are selected from 
\begin{align*}
\ev_{\Usf\binom{\Ksf-1}{\Ssf-1},j \binom{\Ksf-1}{\Ssf-1}+i}, \ \forall j\in [0:\Usf-1], i\in \left[ \binom{\Ksf-1}{\Ssf-1}-\binom{\Ksf-1-\Usf}{\Ssf-1}\right], k\in \Cc_{1,i},
\end{align*}
totally $\Usf \binom{\Ksf-2}{\Ssf-2}$ unit vectors. 
Recall that the users in $\Uc_2$ are denoted by  $\Uc_2(1),\ldots, \Uc_2(\Usf)$, where  $\Uc_2(1)<\cdots<\Uc_2(\Usf)$. 
Our objective on the selection is that the $\Usf \left( \binom{\Ksf-1}{\Ssf-1}-\binom{\Ksf-1-\Usf}{\Ssf-1} \right)$ rows in $\begin{bmatrix}
{\bf S}_{\Uc_2(1)}  \\
\vdots\\
{\bf S}_{\Uc_2(\Usf)}
\end{bmatrix}$
are exactly 
\begin{align*}
\ev_{\Usf\binom{\Ksf-1}{\Ssf-1},j \binom{\Ksf-1}{\Ssf-1}+i}, \ \forall j\in [0:\Usf-1], i\in \left[ \binom{\Ksf-1}{\Ssf-1}-\binom{\Ksf-1-\Usf}{\Ssf-1}\right];
\end{align*}
thus the matrix in~\eqref{eq:decodability matrix} is an identity matrix (with some row permutation) and is full rank. 

The existence of the above selection is equivalent to the following combinatorial problem: \\
{\it (p1). There are $\binom{\Ksf-1}{\Ssf-1}-\binom{\Ksf-1-\Usf}{\Ssf-1}$ urns, where each urn is with the index $\Vc \in \binom{[\Ksf]}{\Ssf}$ where $\Uc_2\cap \Vc \neq \emptyset$ and $u\in \Vc$. There are $\Usf$ colors of balls, where the number of balls for each color with index $k\in \Uc_2$ is $\binom{\Ksf-1}{\Ssf-1}-\binom{\Ksf-1-\Usf}{\Ssf-1}$. A ball with color $k$ can be only put into an urn with index $\Vc$ where $k\in \Vc$. We want to put the balls into the urns such that each urn contains $\Usf$ balls (not necessarily with the different colors). } \\
If there exists a solution for Problem (p1), we can treat each color as one user in our problem and each urn as a set in  $\Vc \in \binom{[\Ksf]}{\Ssf}$ where $\Uc_2\cap \Vc \neq \emptyset$ and $u\in \Vc$. Assume the urn with index $\Vc$ contains $x_1$ balls with color $k_1$, $x_2$ balls with color $k_2$, etc. Then for user $k_1$, we select 
$\ev_{\Usf\binom{\Ksf-1}{\Ssf-1},j \binom{\Ksf-1}{\Ssf-1}+i}$ where $j\in [k_1]$ and $\Cc_{1,i}=\Vc$ and put them into ${\bf S}_{k_1}$; for user $k_2$, we select 
$\ev_{\Usf\binom{\Ksf-1}{\Ssf-1},j \binom{\Ksf-1}{\Ssf-1}+i}$ where $j\in [k_1+1:k_2]$ and $\Cc_{1,i}=\Vc$ and put them into ${\bf S}_{k_2}$, etc. Thus we can see that from the  solution for Problem (p1), we can design the selection of the   coefficients satisfying the matrix in~\eqref{eq:decodability matrix} is full rank. 

At the end of this section, we provide one solution for Problem (p1), which is based the Pascal's triangle
\begin{align}
\binom{\Ksf-1}{\Ssf-1}-\binom{\Ksf-1-\Usf}{\Ssf-1}= \binom{\Ksf-2}{\Ssf-2}+\binom{\Ksf-3}{\Ssf-2} + \cdots+ \binom{\Ksf-1-\Usf}{\Ssf-2}. \label{eq:pascal trian}
\end{align}
Let us focus on the $\binom{\Ksf-1}{\Ssf-1}-\binom{\Ksf-1-\Usf}{\Ssf-1}$ balls in color $k\in \Uc_2$, and put these balls into urns by the following $\Usf$ step:\\
Step $t\in [\Usf]$. We put one ball in color $k$ into each urn with index $\Vc$ where \\$\Vc\in \binom{[\Ksf]\setminus \{\Uc_2(<k+1>_{\Usf}),\Uc_2(<k+2>_{\Usf}),\ldots,\Uc_2(<k+t-1>_{\Usf})\}}{\Ssf}$, $\Uc_2\cap \Vc \neq \emptyset$, and $\{u,k\} \subseteq \Vc$. Thus in this step, we have put $ \binom{\Ksf-1-t}{\Ssf-2}$ balls in color $k$ into urns.\\
Hence, by the Pascal's triangle in~\eqref{eq:pascal trian}, consider all the $\Usf$ steps for the balls in color $k$, we have put all the $\binom{\Ksf-1}{\Ssf-1}-\binom{\Ksf-1-\Usf}{\Ssf-1}$ balls in color $k$ into urns. 

Let us then show that after considering the balls in all colors, each urn has exactly $\Usf$ balls. Consider an urn with index $\Vc\in \binom{[\Ksf]}{\Ssf}$ where $\Uc_2\cap \Vc \neq \emptyset$ and $u\in \Vc$.  Assume that $\Uc_2\cap \Vc= \Bc=\{\Uc_2(i_1),\Uc_2(i_2),\ldots,\Uc_2(i_{|\Bc|})\}$, where $i_1<i_2<\cdots<i_{|\Bc|}$. 
We consider two cases:
\begin{itemize}
\item $|\Bc|=1$. By construction, in each of the $\Usf$ steps for color $\Bc(1)$, we put one ball in color $\Bc(1)$ into the urn with index $\Vc$. Hence, this urn totally contains $\Usf$ balls.
\item $|\Bc|>1$. By construction, for each $s\in [|\Bc|]$,  in each of the first $<i_{<s+1>_{|\Bc|}}-i_s>_{\Usf}$ steps for color $\Uc_2(s)$, we put one ball in color $\Uc_2(s)$ into the urn with index $\Vc$. Hence, considering all $s\in [|\Bc|]$, the number of balls in this urn is 
\begin{align*}
&<i_{2}-i_1>_{\Usf} + <i_{3}-i_2>_{\Usf} +\cdots+ <i_{|\Bc|}-i_{|\Bc|-1}>_{\Usf}+ <i_{1}-i_{|\Bc|}>_{\Usf}=i_{|\Bc|}-i_1+ <i_{1}-i_{|\Bc|}>_{\Usf}\\
& = \Usf.
\end{align*}
\end{itemize}
As a result, we proved that each urn has exactly $\Usf$ balls. Thus the proposed solution is indeed a solution for Problem (p1). Hence, we showed that  $P(\Ac)$ is a non-zero polynomial and proved Lemma~\ref{lem:SZ lemma}.

\bibliographystyle{IEEEtran}
\bibliography{IEEEabrv,IEEEexample}

\end{document}